\documentclass[preprint,12pt,authoryear]{elsarticle}

\usepackage{graphicx}
\usepackage{color}

\textfloatsep = 0.4in \addtocontents{toc}{\vspace{0.4in} \hfill
Page\endgraf} \addtocontents{lof}{\vspace{0.2in} \hspace{0.13in} \
Figure\hfill Page\endgraf} \addtocontents{lot}{\vspace{0.2in}
\hspace{0.13in} \ Table\hfill Page\endgraf}
\usepackage{float}
\usepackage{textcomp,bm}
\usepackage{array}
\usepackage{rotating}
\usepackage{listings}
\usepackage{pdfpages}
\usepackage{pdflscape}
\usepackage{setspace}
\usepackage{mathptmx}
\usepackage{capt-of}
\usepackage[table, svgnames]{xcolor}
\usepackage{colortbl}
\usepackage{graphicx}
\graphicspath{{}}
\usepackage{amssymb, amsmath}
\usepackage{subfig}
\usepackage{wrapfig}
\usepackage{times}
\usepackage{float}
\usepackage{bm}
\usepackage{lipsum}
\usepackage{url}
\usepackage{multirow}
\usepackage{pdfpages,caption}
\usepackage{booktabs}
\usepackage[subfigure, titles]{tocloft}
\usepackage{caption}
\usepackage{datetime}
\usepackage[autostyle]{csquotes}
\usepackage[linktocpage=true]{hyperref}
 \usepackage[margin=1in]{geometry}
\usepackage{algorithm}
\usepackage{cite}
\usepackage{lscape}
\usepackage{mdframed}
\usepackage{array}
\usepackage{tikz}

\usepackage{makecell}
\usepackage{titletoc}
\usepackage{tikz}
\usetikzlibrary{decorations.pathmorphing} % noisy shapes
\usetikzlibrary{fit}% fitting shapes to coordinates
\usetikzlibrary{backgrounds}	% drawing the background after the foreground
\usetikzlibrary{positioning}
\usetikzlibrary{shapes,decorations,arrows,calc,arrows.meta,fit,positioning}
\tikzset{
    -Latex,auto,node distance =1 cm and 1 cm, semithick,
    state/.style ={ellipse, draw, minimum width = 0.7 cm},
    point/.style = {circle, draw, inner sep=0.04cm,fill,node contents={}},
    bidirected/.style={Latex-Latex,dashed},
    el/.style = {inner sep=2pt, align=left, sloped}
     latentnode/.style={draw, minimum width=5mm, shape=circle, ultra thick, black},
  dagconn/.style={arrows=->, black, thick},
  plate/.style={draw, shape=rectangle, rounded corners=0.5ex, thick, minimum width=3.1cm, text width=3.1cm,inner sep=10pt, inner ysep=10pt, 
    label={[xshift=-14pt,yshift=14pt]south east:#1}}
}

\usepackage[nottoc]{tocbibind}
\IfFileExists{upquote.sty}{\usepackage{upquote}}{}

\begin{document}
\pagenumbering{alph}
\begin{frontmatter}

%% Title, authors and addresses

%% use the tnoteref command within \title for footnotes;
%% use the tnotetext command for theassociated footnote;
%% use the fnref command within \author or \affiliation for footnotes;
%% use the fntext command for theassociated footnote;
%% use the corref command within \author for corresponding author footnotes;
%% use the cortext command for theassociated footnote;
%% use the ead command for the email address,
%% and the form \ead[url] for the home page:
%% \title{Title\tnoteref{label1}}
%% \tnotetext[label1]{}
%% \author{Name\corref{cor1}\fnref{label2}}
%% \ead{email address}
%% \ead[url]{home page}
%% \fntext[label2]{}
%% \cortext[cor1]{}
%% \affiliation{organization={},
%%            addressline={}, 
%%            city={},
%%            postcode={}, 
%%            state={},
%%            country={}}
%% \fntext[label3]{}

\title{A Bayesian Spatial Berkson error approach to estimate small area opioid mortality rates accounting for population-at-risk uncertainty}

\author[1]{Emily N. Peterson\corref{cor1}}
\ead{emily.nancy.peterson@emory.edu}
\affiliation[1]{organization={Department of Biostatistics and Bioinformatics, Emory Rollibs School of Public Health},
            city={Atlanta},
            state={GA},
            country={USA}}

\cortext[cor1]{Corresponding Author}

\author[2]{Rachel C. Nethery}
\author[2]{Jarvis T. Chen}
\author[4]{Loni P. Tabb}
\author[2]{Brent A. Coull}
\author[3]{Frederic B. Piel}
\author[1]{Lance A. Waller}

\affiliation[2]{organization={Department of Biostatistics, Harvard TH Chan School of Public Health},
            city={Boston},
            state={MA},
            country={USA}}
\affiliation[3]{organization={Department of Epidemiology and Biostatistics, School of Public Health, Imperial College London},
            city={London},
            country={UK}}
\affiliation[4]{organization={Department of Epidemiology and Biostatistics, Dornsife School of Public Health, Drexel University},
            city={Philadelphia},
            state = {PA},
            country={USA}}

\begin{abstract}
Monitoring small-area geographical population trends in opioid mortality has large scale implications to informing preventative resource allocation. A common approach to obtain small area estimates of opioid mortality is to use a standard disease mapping approach in which population-at-risk estimates are treated as fixed and known. Assuming fixed populations ignores the uncertainty surrounding small area population estimates, which may bias risk estimates and under-estimate their associated uncertainties. We present a Bayesian Spatial Berkson Error (BSBE) model to incorporate population-at-risk uncertainty within a disease mapping model. We compare the BSBE approach to the naive (treating denominators as fixed) using simulation studies to illustrate potential bias resulting from this assumption. We show the application of the BSBE model to obtain 2020 opioid mortality risk estimates for 159 counties in GA accounting for population-at-risk uncertainty. Utilizing our proposed approach will help to inform interventions in opioid related public health responses, policies, and resource allocation. Additionally, we provide a general framework to improve in the estimation and mapping of health indicators.

\end{abstract}

\begin{keyword}
%% keywords here, in the form: keyword \sep keyword
Spatial uncertainty \sep Disease mapping \sep Berkson measurement error \sep Opioid mortality trends 
%% PACS codes here, in the form: \PACS code \sep code

%% MSC codes here, in the form: \MSC code \sep code
%% or \MSC[2008] code \sep code (2000 is the default)

\end{keyword}
\end{frontmatter}
%%%%%%%%%%%%%%%%%%%%%%%%%%%%%%%%%%%%%%%%%%%%%%%%%%%%%%%%%%%%%%%%%%%%%%%%%%%%%%%%

\pagenumbering{roman} \setcounter{page}{1}
\captionsetup{font=small,skip=0pt}

\tableofcontents

%%%%%%%%%%%%%%%%%%%%%%%%%%%%%%%%%%%%%%%%%%%%%%%%%%%%%%%%%%%%%%%%%%%%%%%%%%%%%%%%
\begingroup
\setlength{\parskip}{1\baselineskip}
\newpage
\endgroup

%%%%%%%%%%%%%%%%%%%%%%%%%%%%%%%%%%%%%%%%%%%%%%%%%%%%%%%%%%%%%%%%%%%%%%%%%%%%%%%%
\normalsize

\pagenumbering{arabic}
\setcounter{page}{1}
\singlespacing

 \section{Introduction}
The standard small area disease mapping methods used to assess spatial distributions of disease and/or mortality risks do not account for potential bias and uncertainty associated with population-at-risk estimates. As such, resulting small area risk estimates of disease/mortality may be inaccurate.
% Small area estimation operates at the interface of design-based and model-based statistical inference, where historical motivation involved the development of statistically robust, model-based estimates for population subgroups with too few observations for accurate survey design-based estimates. 
The generic disease mapping framework consists of a Poisson regression of the observed incidence counts adjusting for local covariate values and the size of the population-at-risk (referred to as the offset), which is often treated as fixed and known \citep{waller_gotway, jon_wakefield_disease_2007}. Population-at-risk values are commonly derived from U.S. Census data products (Decennial Census, Population Estimation Program, and American Community Survey) that report small area population counts \citep{census, us_census_bureau_understanding_2018, pep}. Additionally, recent spatial mapping innovations have produced alternative demographic data sources such as WorldPop which produce high spatial resolution data on human population distributions to address current limitations in national censuses and health surveillance systems \citep{worldpop}. Although each source reports population data for the same set of small areas, there are important distinctions in data collection and processing methodologies and availability, which yield population-at-risk estimates that suffer from varying types and degrees of error. To accurately capture small area disease/mortality risks, we must incorporate the uncertainty associated with offset estimates within the disease mapping model while also accounting for the source of the reported denominator data, and its respective degree of error.\\
 % As such, small area disease or mortality rates can differ depending on which source of population data are used as denominators (offsets) (cite Rachel). 
% Uncertainty associated with population-at-risk estimates is commonly ignored in standard small area estimation, which may bias local estimates of disease/mortality risk and under-estimate associated uncertainties. Consequently, 

\noindent
In the United States, annual small area (county and census-tract level) population counts are published by the United States Census Bureau (USCB) in the form of the decennial census, intercensal population projections (PEP), and the American Community Survey (ACS) multi-year estimates \citep{census, us_census_bureau_understanding_2018, pep}.  The decennial census is a cross-sectional comprehensive survey mandated every 10 years to count the entire U.S. population, which is accomplished through multiple modes of collection. Census counts do not suffer from sampling error, but do suffer from forms of non-sampling measurement error (i.e., duplications, erroneous errors, and omissions) \citep{ census, CC, us2004accuracy,  nonsamp}. PEP reported intercensal population estimates are derived from a cohort component model, which uses the last decennial census as a base population, and projects population estimates forward using births, deaths, and net migrations \citep{pep,demog}. As such, PEP-reported population estimates suffer from unknown non-sampling errors including census-related errors and errors associated with birth, death, and migration data. USCB formally recommends the use of PEP or decennial counts as population estimates, however, PEP reported population counts are not available for geographies smaller than county, i.e., census tracts and block groups. The ACS is a complex rolling sample survey conducted annually, which collects 3.5 million independent samples of data nationally (approximately 2.5\% of the population) for each year within a 5-year time interval. ACS reports small area (county, census tract, block groups) population counts using data sampled over the 5 year time interval, referred to as multiyear estimates. Additionally, the ACS reports associated margins of error, which quantify the uncertainty (variability) due to sampling error across multiple years \citep{us_census_bureau_american_2014-2, us_census_bureau_american_2014-1, us_census_bureau_understanding_2009, us_census_bureau_understanding_2018}.\\

\noindent
Private companies and academic groups have begun to produce high resolution gridded population estimates based on machine learning (ML) models that often combine census, remote sensing, land use, and other information to estimate population counts at smaller geographies in near real time. One of the most popular products of this nature is WorldPop (WP), which utilizes an open-source algorithm and provides yearly global high resolution gridded population estimates. Advantages of WP include its near real-time capability (available for the current year) and high spatial resolution \citep{worldpop, nethery}. In summary, WP uses a combination of available, remotely-sensed and geospatial datasets (i.e., settlement locations, settlement extents, land cover, roads, building maps, health facility locations, satellite nightlights, vegetation, topography, refugee camps) incorporated within a random forest model to generate gridded predictions of population density at $\sim 100m$ spatial resolution across the globe. Gridded estimates are weighted and aggregated to produce small area to regional level estimates of population size and other demographic indicators \citep{worldpop, stevens, tatem2}. High spatial resolution population estimates derived from the ML algorithms address some of the limitations of official population statistics, however, do not adhere to the same validation and control measures as official population counts. Additionally, these algorithms are often trained on census reported population counts and therefore inherit the bias present in census reported counts. These notable and important differences in data collection, availability, and validation methodologies highlight the need for data-source specific mechanisms of incorporating offset related error within a disease mapping model. 
\\

\noindent
There is an abundance of rich literature on measurement error methods to account for error in covariates  using both classical and Berkson error methods \citep{gustafson,  raymond_j_carroll_measurement_2006}. \citet{raymond_j_carroll_measurement_2006} highlighted the use of measurement error corrections in the context of Bayesian epidemiological studies. Refer to \ref{sec:berkson} for a summary of the Bayesian Berkson error approaches. However, relatively little literature addresses measurement error in the context of spatial data \citep{huque, huque2, li, zhang, josey}, and there is an absence of literature addressing measurement error associated with population-at-risk values. \citet{li} quantified the impact of ignoring measurement error on spatial data analysis, and showed naive estimators of regression coefficients are attenuated towards the null while variance components are inflated and biases are related to spatial dependence parameters. One previous study by \citet{zhang} assessed small area risk estimates of fatal car crashes accounting for offset related errors associated with using a proxy denominator variable combining Berkson and disease mapping methodologies. \citet{peterson} developed a Bayesian hierarchical small area population (BPop) model to produce county-level population estimates fusing information across USCB data products (census, ACS, PEP) while accounting for data source specific errors within the model framework. To our knowledge, no study has addressed how to incorporate offset uncertainty within a larger disease mapping model across multiple types of denominator data sources and when population-at-risk associated errors are unknown.\\

\noindent
To accurately estimate small area risks accounting for offset uncertainty we present a Bayesian Hierarchical Spatial Berkson Error (BSBE) model, which fuses modeling approaches of the  Besag-York-Mollie (BYM) disease mapping model \citep{bym, andrea_riebler_intuitive_2016} and Berkson measurement error methods \citep{raymond_j_carroll_measurement_2006}, in which we assume the true population-at-risk (offset) is unknown, and is derived as a function of the reported population size plus associated error \citep{raymond_j_carroll_measurement_2006, gustafson, huque, huque2}. We assess our proposed approach across three denominator data sources (PEP, ACS, and WorldPop) to illustrate incorporation of  different data source error specific mechanisms across varying degrees of error and availability of information. In the case of PEP intercensal and WP population counts, there is no available information on offset related errors. As such, we incorporate offset-uncertainty using a model-based hierarchical approach. In contrast, ACS estimates provide direct information on the degree of sampling error associated with population counts, but also do not report on non-sampling errors. Lastly, WP does not report model-based errors associated with their gridded estimates, but also is not required to be consistent with official statistics data. Outside of these respective sources, we note the BSBE model has broad applicability across data sources, geographies, demographic groups, and health indicators of interest. \\

\noindent
We assess and illustrate model results using both: (1) A simulation study to compare model performance between our proposed BSBE approach and the naive approach of assuming population-at-risk values are fixed, and (2) An application of the BSBE approach to obtain county age-stratified 2020 estimates of opioid-related mortality risks and associated uncertainties for the 159 counties in the state of Georgia. The United States is in the midst of a continuing public health crisis due to opioid misuse and overdoses. Accurate estimation of small area risks of opioid mortality is essential to properly target harm reduction resources in effective and reliable ways. Disease mapping models have been widely used to assess small area trends in opioid mortality rates and commonly use USCB small area data products such as census, PEP, or ACS population counts as denominator data \citep{kline, klinespat, klinespat2, rossen}. A limitation of the current research in opioid mortality estimation is the lack of accounting for errors associated with these denominators. We apply the BSBE approach to estimate small area (county-level) opioid mortality trends in GA accounting for data source specific uncertainty, and to highlight notable differences in small area results between our approach and the naive approach, which results in identification of different high need areas.  \\

\noindent
The paper is organized as follows: Section \ref{sec:desc} describes population-at-risk uncertainty across the different denominator sources (PEP, ACS, and WP). Section \ref{sec:summary} summarizes the BSBE approach to incorporate population-at-risk uncertainty within a disease mapping model. Section \ref{sec:sim} outlines the process used to assess the impact of offset uncertainty on small area risk estimates using simulation studies and respective findings. Section \ref{sec:opapp} demonstrates the application of the BSBE approach to obtain age-stratified small area opioid-related mortality risk estimates. Lastly, Section \ref{sec:discussion} discusses summary of findings, limitations, and implications of our study. 

\section{Methods}\label{sec:methods}

\subsection{Description of population-at-risk uncertainty}\label{sec:desc}
\begin{figure}[H]
\center
\includegraphics[width=0.95\textwidth]{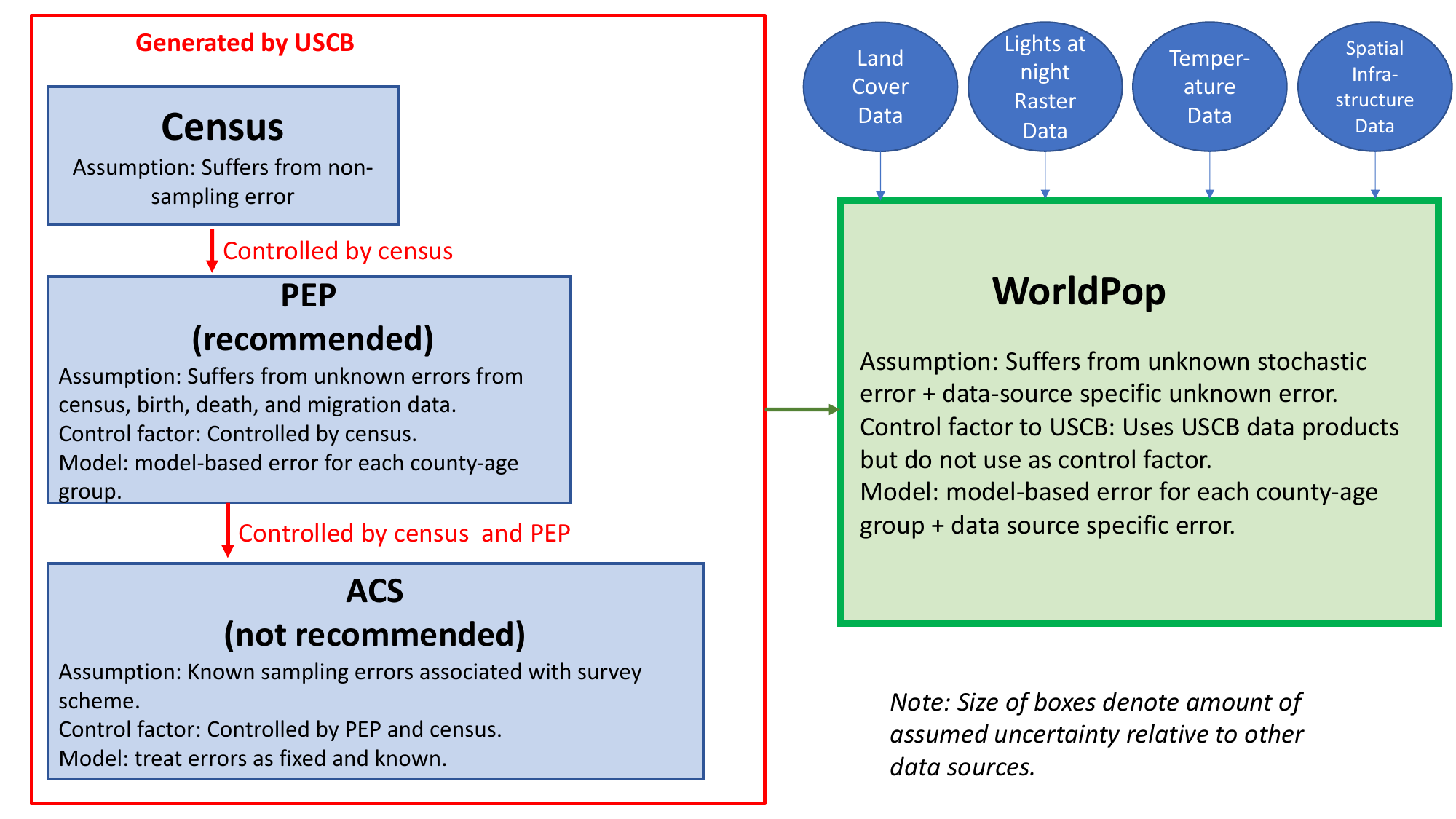}
\caption{Diagram of the source and structure of uncertainties assumed by the BSBE model broken down by data type. Size of box denotes the amount of uncertainty imposed relative to the other data sources. Blue boxes refer to population data sources generated by USCB. Green boxes refer to alternative data sources (WP). The arrow from USCB to WP indicates the use of USCB data by WP. (Recommended) refers data sources recommended by USCB to be used as population-at-risk denominator data. WP data sources shown in blue circles.}
\label{fig:poperr}
\end{figure}

\noindent
To motivate our model assumptions, we reference previous work comparing total ACS reported uncertainties with those of the 2000 decennial census long form data, which showed that ACS margins of error were, on average 75\% larger than those corresponding to the 2000 decennial census (due to the smaller sample sizes), representing a considerable increase in the uncertainty of values central to disease mapping \citep{seth_e_spielman_reducing_2015}. Figure \ref{fig:poperr} illustrates the hierarchy of error imposed within our Berkson error approach based on model assumptions. We assume U.S. decennial census data (top blue box) suffers from unknown non-sampling error, but that error is minimal compared to other data sources due to the comprehensive nature of the U.S. decennial census. PEP intercensal projections (middle blue box) use decennial census data as their base population counts, therefore, we assume that PEP also suffers from unknown error from census, birth, death, and migration data, but this error is reduced compared to ACS population estimates, i.e., PEP non-sampling errors are less than ACS sampling errors, keeping consistent with the above findings which confirms ACS sampling error to be substantially larger than the census-related non-sampling error. As such, we assume ACS (bottom blue box) suffers from the highest degree of error within USCB data sources due to complex sampling and weighting structures used to obtain population estimates. WP population estimates suffer from unknown stochastic errors, which can be attributed to the use of multiple data sources (land cover data, temperature data, census data, etc.), and model-based errors. As such, we assume that the total variance associated with WP reported population-at-risk estimates can be broken down into two components: (1) a model-based stochastic error term which captures the uncertainty induced by a model-based approach, and (2) a data source specific error term capturing the error associated with the use of multiple data sources. 

\begin{figure}[H]
\center
\includegraphics[width=\textwidth]{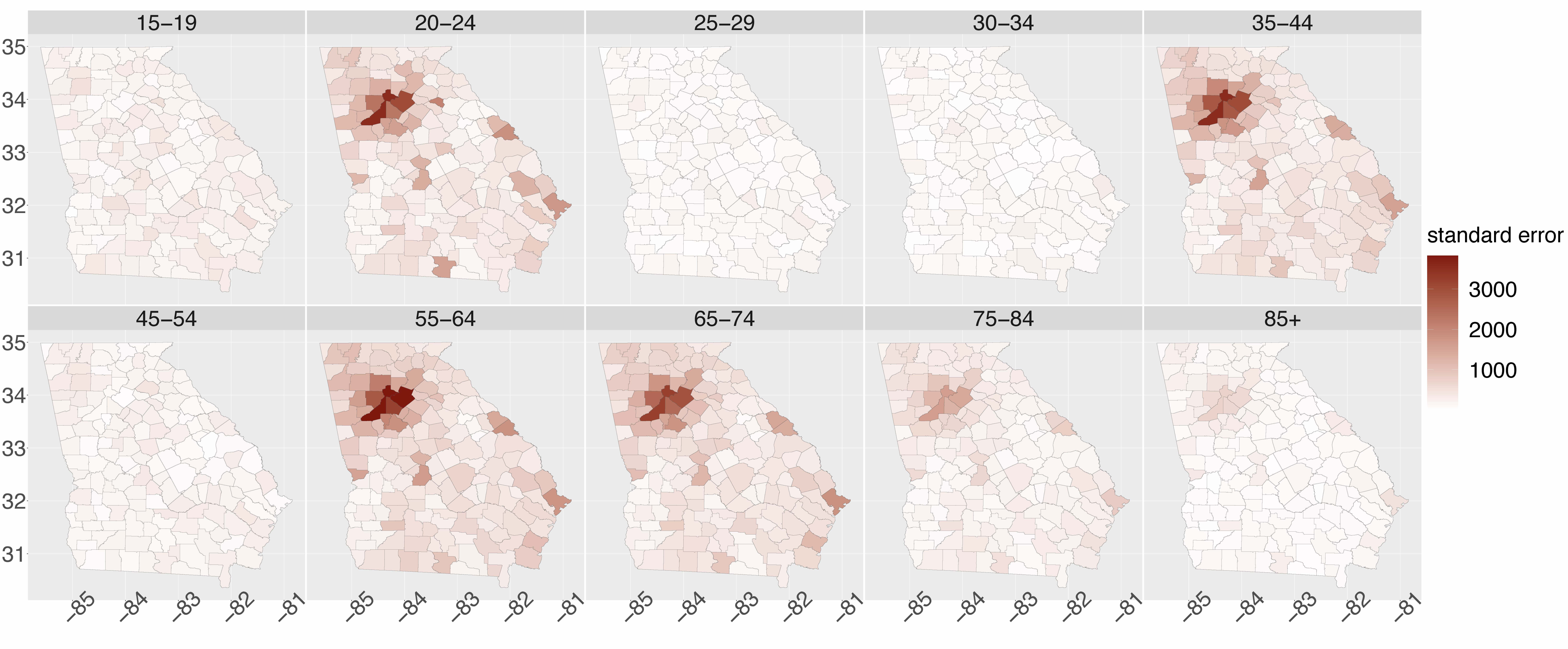}
\caption{Mapped ACS reported standard errors across 159 counties in Georgia, stratified by 5 year age group intervals. Scale ranges from small offset error (white) to large offset error (dark red).}
\label{fig:acserr}
\end{figure}

\noindent
Importantly, one critical aspect of ACS related errors is that uncertainty is not uniform across geographic areas with significant regional variation in precision, i.e., some places have more precise data \citep{spielman_s_patterns_2014, starsinic}. For those population data sources with unobserved offset errors (PEP and WP), area-specific errors must be modeled hierarchically. The model structure assumed for unobserved offset-related errors can be motivated by the ACS reported standard errors, in which we can ascertain spatial and demographic patterns of population-at-risk uncertainty. Figure \ref{fig:acserr} maps the ACS reported standard errors associated with their population estimates by county and age-group. The mapped ACS reported standard errors illustrate that there exists varying spatial patterns of offset related errors across the different age-groups. Notably, the spatial patterns of error show higher standard errors in more heterogenous counties, i.e., DeKalb, Fulton, and Gwinnett counties which have the largest populations in GA and are spatially in close proximity to each other. More rural and homogenous counties show lower levels of offset related error. For example, the largest standard error $(sd = 3,823)$ is associated with Gwinnett County for the 55-64 age group. Conversely, the smallest standard error $(sd = 9.72)$ is associated with Clarke County for the 25-29 age group. We also assessed spatial patterns of relative error (sd/population size), which showed larger counties (i.e., counties with larger populations) had smaller relative error compared to counties with smaller populations due to larger sample sizes. Smaller populations suffering from higher degrees of relative error result in higher uncertainty in relative risk estimates of disease/mortality. Refer to \ref{sec:relerr} for mapped relative errors across counties in GA. These results highlights the need to account for spatial variation of errors across different geographies and populations.

\subsection{Summary of BSBE}\label{sec:summary}
\noindent
Our Bayesian Spatial Berkson Error (BSBE) model framework aims to estimate small-area disease/mortality rates accounting for uncertainties associated with small-area population size estimates across the above described data types. In our model set-up we focus on estimating county-age-group specific rates, but note our approach could be applied to various spatial resolutions and socio-demographic sub-groups. The BSBE model can be broken down into three main components: (1) the data model for observed disease counts, (2) the process model for latent relative risks incorporating an alternative BYM parametrization, and (3) the process model for the true log-transformed offsets. The main features of the model are as follows:

\begin{enumerate}
 \item The data model (defining the likelihood function) consists of modeling observed county and age-group specific disease/death counts $y_{c,a}$ using a Poisson assumption and is further detailed in Section \ref{sec:dm}. 
 \item The log-relative risk for county $c$ age-group $a$ ($\mu_{c,a}$) is modeled as a function of county-age-group specific covariates $X_{c,a}$ and spatial random effects $(\theta_c + \phi_c)$. This is further described in Section \ref{sec:process}. 
 % \item We propose an alternative parametrization of the spatial BYM parametrization used in standard disease mapping models, in which we combine area-level non-spatial random effects. Due to this reparametrization, the variance of the area-specific unstructured random effect $\kappa_{i,r}$ is the sum of the unstructured variance and the reported sampling error associated with population estimates. This is further described in Section XX.
\item True and unknown log-transformed population-at-risk, denoted $log(\gamma_{c,a})$, are modeled using a Berkson error model, which assumes the truth is normally distributed around the observed population data, and is further detailed in Section \ref{sec:poperr}.
  \end{enumerate}

\subsubsection{Data model for observed disease counts}\label{sec:dm}
\noindent
In a standard disease mapping model we assume that the relation between the observed disease/mortality counts and the relative risk parameter (i.e., the data model for the observed cases) is given by a Poisson distribution shown in Eq. \ref{eq:dm1}. We denote observed counts for county $c$ age-group $a$ (labeled $[c,a]$ for ease of readability) as $y_{c,a}$. In a standard disease mapping model, a Poisson data generating assumption is assumed, i.e, $y_{c,a}|E_{c,a}, \mu_{c,a} \sim Poisson (E_{c,a} \cdot exp(\mu_{c,a}))$. The expectation is written as a function of the expected count $E_{c,a}$ defined as the product of a reference rate $R$ and the observed population-at-risk $n_{c,a}$. The second term in the expectation is the log relative risk for a given county-age-group denoted $\mu_{c,a}$. Both the reference rate and the population-at-risk are treated as fixed and known. The global rate is derived across populations and areas (from a much larger sample size) than are the local estimates, and so suffers from substantially less relative uncertainty and for simplicity we treat as fixed in our model approach, i.e., $R=\sum_i y_i/\sum_i N_i$ across all populations $i$. We incorporate uncertainty associated with population-at-risk estimates by replacing the observed transformed population-at-risk $n_{c,a}$ with an unknown true population-at-risk $\gamma_{c,a}$.  We re-parametrize the model to include the log-transformed offset $log(E_{c,a}) =  log(R) + log(\gamma_{c,a})$ in the exponential term shown in Eq. \ref{eq:proc}.

\begin{align}
y_{c,a}|E_{c,a},\mu_{c,a} &\sim Poisson(exp(\omega_{c,a}))\label{eq:dm1}\\
\omega_{c,a} &=\mu_{c,a} + log(R) + log(\gamma_{c,a})\label{eq:proc}\\
\mu_{c,a} &= X_{c,a}'\beta  + \delta(\sqrt{\rho}\cdot\theta_c^* + \sqrt{1-\rho}\cdot\phi_c^*)\\
\theta^*_c &\sim N(0, 1)\nonumber\\
\phi^*_c & \sim ICAR(1)\nonumber\\
\beta_k & \sim N(0, 5^2)\nonumber\\
\rho & \sim Unif(0,1)\nonumber\\
log(\delta) & \sim Gamma(0, 0.001)\nonumber
\end{align}

\subsubsection{Process model for latent relative risks}\label{sec:process}
\noindent
We capture the association between log relative risk $\mu_{c,a}$ and the covariates through the vector of global covariate coefficients $\bm{\beta}$, which are modeled using non-informative $N(0, 5^2)$ priors. The spatially structured and unstructured random effects are often parametrized using the sum of the components $(\theta_c + \phi_c)$ \citep{bym}. Riebler et al. (2016) proposed an alternative parameterization (BYM2) in which the spatial random effects are scaled to have an approximate variance of one, i.e, $Var(\theta^*) \approx Var(\phi^*) \approx 1$ to ascertain how much variance is attributed to the spatial autocorrelation random effect versus the unstructured random term \citep{andrea_riebler_intuitive_2016}. Let $\theta^*$ and $\phi^*$ denote the scaled spatially structured and unstructured effects. The BYM2 parametrization places a single precision (scale) parameter $\delta$ on the combined components, and a mixing parameter $\rho$ determining the amount of variation attributed to the spatially structured effect. The spatially structured term, $\theta_c^*$ is modeled with a intrinsic conditional autoregressive (ICAR) prior, and the spatially unstructured term $\phi_c^*$ is modeled using a standard normal N(0,1) prior. Vague priors are placed on global hyper-parameters $(\beta, \rho, \delta)$ shown in Eq. \ref{eq:proc}. 
    
% \noindent
% The spatial random effect $\theta_c$ captures spatial interactions between pairs of areas $c$ and $j$, which can be modeled conditionally  as a normal random variable. In the full conditional distribution, each $\phi_c$ is conditional on the sum of the weighted values of its neighbors $(w_{cj}\phi_c)$ and has unknown variance.
% \begin{equation}
%     \phi_c|\phi_j, j\neq c \sim N(\sum_{j=1}^n w_{cj} \phi_j, \sigma^2)
% \end{equation}
% Alternatively, we can write the vector of area-specific effects as a multivariate normal, i.e., $\phi \sim N(0, Q^{-1})$. The variance of  $\phi$ is specified as a precision matrix $Q$, which is the inverse of the covariance matrix $\Sigma$, i.e., $\Sigma = Q^{-1}$. 
% \begin{equation}
%     Q = D(I-\alpha A)
% \end{equation}
% where $D$ denotes the diagonal neighbor matrix, $I$ the identity matrix, and $\alpha$ controls the amount of spatial dependence scaling the matrix $A$. In the intrinsinc conditional autoregressive models, $\alpha$ is set equal to 1, such that $\phi_c$ is normally distributed with a mean equal to the average of its neighbors. Its variance decreases as the number of neighbors denoted $d_c$ increases. The conditional specific is given by,
% \begin{equation}
%     p(\phi_c|\phi_{c\sim j}) = N\left(\frac{\sum_{c\sim j} \phi_c}{d_c}, \frac{\sigma_c^2}{d_c}\right)
% \end{equation}

% We use a standard $Normal(0,100)$ prior distribution assumption for the covariate effects $(\bm{\beta}, \pi)$, and log-$\pi$ distributions for the variance components as the default prior defined in $R-INLA$.

\subsubsection{Incorporation of population-at-risk uncertainty across data types}\label{sec:poperr}
The model for the unobserved true log-transformed population-at-risk $log(\gamma_{c,a})$ uses a Berkson error approach in which we assume the truth is centered around the observed value plus some error \footnote{Refer to \ref{sec:berkson} for a brief summary of the Berkson error method.}. We account for varying degrees of error across the different data sources by incorporating data source specific error models. Below we describe in further detail each data source specific error model for PEP, ACS, and WP.

\noindent
\textit{PEP:} We account for the unknown PEP related uncertainty using the Berkson error approach where we assume the true unknown log offset for each $[c,a]$, denoted $log(\gamma_{c,a})$, is normally distributed centered around the observed population denoted $log\left(n_{c,a}^{(PEP)}\right)$, with an unknown county-age-group specific variance $\sigma^{2}_{c,a}$. The model for the standard error terms is motivated by findings in Figure \ref{fig:acserr} which depicts spatial relationships among ACS reported errors. The unknown standard error terms are modeled hierarchically in which we inform unknown age-group specific log-transformed standard errors denoted $log(\bm{\sigma}_{a}) = log(\sigma_{1:N,a})$ using an ICAR prior imposing spatial correlation of population-at-risk (offset) uncertainty across $N$ total counties. As such, we are assuming that standard errors of neighboring areas are similar. Let $D$ be the diagonal matrix of number of neighboring counties for a given county, i.e, $d_{ii}$ is the number of neighbors for county $i$. The adjacency matrix $W$ determines neighborhood structure, in which entries $\{i,i\}$ are zero, and off diagonal elements are 1 if counties $i$ and $j$ are neighbors. We assume that within in each age strata, errors are spatially structured. 
% The standard error terms are modeled hierarchically using a truncated normal prior as recommended by Gelman et al. \citep{andrew_gelman_prior_2006}. This has the desired effects of inducing smaller standard errors compared to alternative data sources, i.e., weakly informative prior with a distribution centered at 0, and bounded below by 0 with small deviation of 1.
\begin{align}
    log(\gamma_{c,a}) &\sim N\left(log\left(n_{c,a}^{(PEP)}\right), \sigma^2_{c,a}\right)\label{eq:pop_pep}\\
   log(\bm{\sigma}_{a})& \sim N(0, [\omega(D-W)]^{-1}) \nonumber
\end{align}

\noindent
\textit{ACS:}
In the case of ACS data, population estimates and associated margins of error (MOEs) are reported for each $[c,a]$ which captures the variability due to sampling error. Standard deviation for each stratified population is given by $sd = MOE/1.645$. We obtain the observed ACS reported sampling errors related to log-transformed population estimates, denoted $s_{c,a}$ using the delta method.
% sampling approach in which we sample 1000 values from a normal distribution centered around the ACS reported population estimate and standard error, $n^{(s)}_{c,a} \sim N\left(n_{c,a}^{(ACS)}, sd^{(ACS)}_{c,a}\right), \text{ for } s=1,...,1000$. We transform the sampled values to the log-scale to obtain their measure of error, $s^2_{c,a} = var(log(n^{(s)}_{c,a}))$. 
We assume the true unknown log offset $ log(\gamma_{c,a})$ is normally distributed centered around the observed population $log\left(n_{c,a}^{(ACS)}\right)$, with a known variance $s^2_{c,a}$ derived from ACS standard errors.
\begin{align}\label{eq:pop_acs}
    log(\gamma_{c,a}) &\sim N\left(log\left(n_{c,a}^{(ACS)}\right), s^2_{c,a}\right)
\end{align}

\noindent
\textit{WorldPop:}
In comparison to USCB data sources, we assume a higher degree of variability due to three characteristics of WP derived population estimates: (1) WP estimates are not calibrated to official statistics, (2) WP estimates are drawn from multiple data sources, and (3) WP estimates are produced at fine spatial resolutions, which suffer from higher degrees of relative uncertainty compared to those aggregated to larger geographic areas, i.e., counties or states. To impose increased variation, we assume the true unknown log offset $ log(\gamma_{c,a})$ is normally distributed centered around the observed population $log\left(n_{c,a}^{(WP)}\right)$, and the unknown variance is equal to a model-based stochastic area-group specific error $\sigma^2_{c,a}$ plus a data-source specific error denoted $\sigma_{(WP)}^2$. The county-age-group stochastic errors are modeled using the ICAR prior given in Eq. \ref{eq:pop_pep}. The data-source specific error captures the additional variability attributed to WP model-based methods and is modeled hierarchically using truncated normal distribution recommended by \citet{andrew_gelman_prior_2006}, which imposes a lower bound of 0, denoted $N_{[0,]}$. This gives the desired effect of having the lower bound of the distribution at 0, while imposing a large variance of 10. 
\begin{align}\label{eq:pop_wp}
    log(\gamma_{c,a}) &\sim N(log(n_{c,a}^{(WP)}), \sigma^2_{c,a} + \sigma_{(WP)}^2)\\
     \sigma_{(WP)}& \sim N_{[0,]}(0, 10)\nonumber
\end{align}

\subsection{Computation}
\noindent
We extract ACS, decennial, and PEP reported population estimates, and ACS margins of error, for 159 counties in Georgia, years 2010-2021, using the tidycensus package \citep{walker}. For model processing and output, a Markov Chain Monte Carlo (MCMC) algorithm samples from the posterior distribution of the parameters via the software $Nimble$ \citep{nimble}. Eight parallel chains were run with a total of 80,000 iterations in each chain. Of these, the first 20,000 iterations in each chain are discarded so the resulting chains contain 60,000 samples. Additionally, we thinned the samples to retain every 10th iteration after burn-in. Thus for each parameters there was a total 48,000 saved posterior samples. Standard diagnostic checks using traceplots were used to check convergence \citep{plummer_jags_2017,gelman_inference_1992,aki_vehtari_rank-normalization_2021, su_r2jags_2020}.

\section{Simulation study}\label{sec:sim}
\noindent
Our aim is to assess the impact (bias) of offset uncertainty on resulting small area-group specific disease or mortality rates, and associated uncertainties. To assess the extent of bias introduced, we perform a simulation study, which compares two approaches for modeling population data (naive versus BSBE) across the different data types. The naive model is defined using a standard disease mapping model in which population-at-risk is treated as fixed and known. We first introduce the data generating process in Section \ref{sec:dg}, in which we generate outcome data for 159 counties in Georgia across 10- 5-year age-groups based on the Berkson model assumption. Following the data generation stage, we fit the naive and BSBE models across the three data types, and compare measures of model performance summarized in  Section \ref{sec:simres}.  
\bigskip

\subsection{Data Generating Process}\label{sec:dg}
\noindent
We generated (simulated) opioid mortality counts based on the Berkson model assuming a Poisson data generating assumption shown in Eq. \ref{eq:dm1}. Using * to denote known simulated quantities, we set the reference rate to be $R^*=0.001$ imposing small counts across counties and age groups. We use standard normal distributions $X\sim N(0,1)$ to derive two covariates denoted $X^*_{1,c,a}$ and $X^*_{2,c,a}$. We fixed covariate coefficient parameters to be $\beta_0^* = 0.001, \beta_1^* = 0.02, \beta_2^* = 0.01$. To impose noise in the county-age-group populations-at-risk we use a sampling approach in which we draw from a normal distribution parametrized using the ACS reported population size and associated variance. The resulting simulated population data inherit the spatial structure present in both ACS reported population sizes and standard errors. We summarize the data generation procedure in Figure \ref{fig:box1}.

\begin{figure}[H]
\begin{mdframed}
\textbf{Summary of data generation settings}
Note: we differentiate generated components, versus estimates obtained via model fitting, using $(*)$.
\begin{itemize}
\item Step I: Fix global parameter and covariate values:
      \begin{itemize}
      \item Fixed covariate coefficients: $\beta_0^* = 0.001, \beta_1^* = 0.02, \beta_2^* = 0.01$
      \item Fixed reference rate: $R^* = 0.001$ based on global set of observed data.
      \item Generate standardized covariate values: $X^*_{1,c,a} \sim N(0,1) $ and $X^*_{2, c, a} \sim N(0,1)$.
      \end{itemize}
      
\item Step II: Generate population-at-risk with stochastic error.
\begin{itemize}
    \item We draw from a normal distribution centered around the ACS reported population count $n_{c,a}^{(ACS)}$, with variance set to the ACS reported variance $s_{c,a}^{(ACS)^2}$.
\end{itemize}
\[n^*_{c,a} \sim N\left(n_{c,a}^{(ACS)}, s_{c,a}^{(ACS)^2}\right)\]

\item Step III: Generate county-age-group specific log relative risk values, $\Omega^*_{c, a}$ based on fixed values from step 1. 
\[\Omega^{*}_{c,a} = X^*_{c,a}\beta^*\]

\item Step IV: Generate county-age group counts $y^*_{c,a}$,  based on settings in steps 1-3, using Poisson data generating assumption.
\[y^*_{c,a} \sim Poisson(R^* \cdot n^{*}_{c,a} \cdot exp(\Omega^*_{c,a}))\]

\item Step V: Repeat steps I to IV for 50 iterations based on these fixed values yielding 50 simulated data sets.
\end{itemize}
\end{mdframed}
\caption{Summary of data generation approach. Describes steps to simulate county-age group-specific disease counts for 159 counties in Georgia across 10 age-group intervals.}
\label{fig:box1}
\end{figure}

\subsection{Model Fitting}
\noindent
We fit the BSBE and naive models to our 50 simulated data sets to obtain model results. Using the BSBE approach, we fit the model defined in Eqs. \ref{eq:dm1}-\ref{eq:proc} across the three data sources. To account for population-at-risk uncertainty, we model the unknown log-transformed true population counts $log(\gamma_{c,a})$ using the population-at-risk models defined in Eqs. \ref{eq:pop_pep}-\ref{eq:pop_wp} across the different data sources. Note that in the case of ACS, we use the ACS reported population counts directly, i.e., $n_{c,a}^{(ACS)}$. In the case of the naive model, we directly plug in denominator data reported by PEP, ACS, and WP thereby treating the population-at-risk as known.  

\subsection{Measures to compare model performance}\label{sec:comp}
\begin{figure}[H]
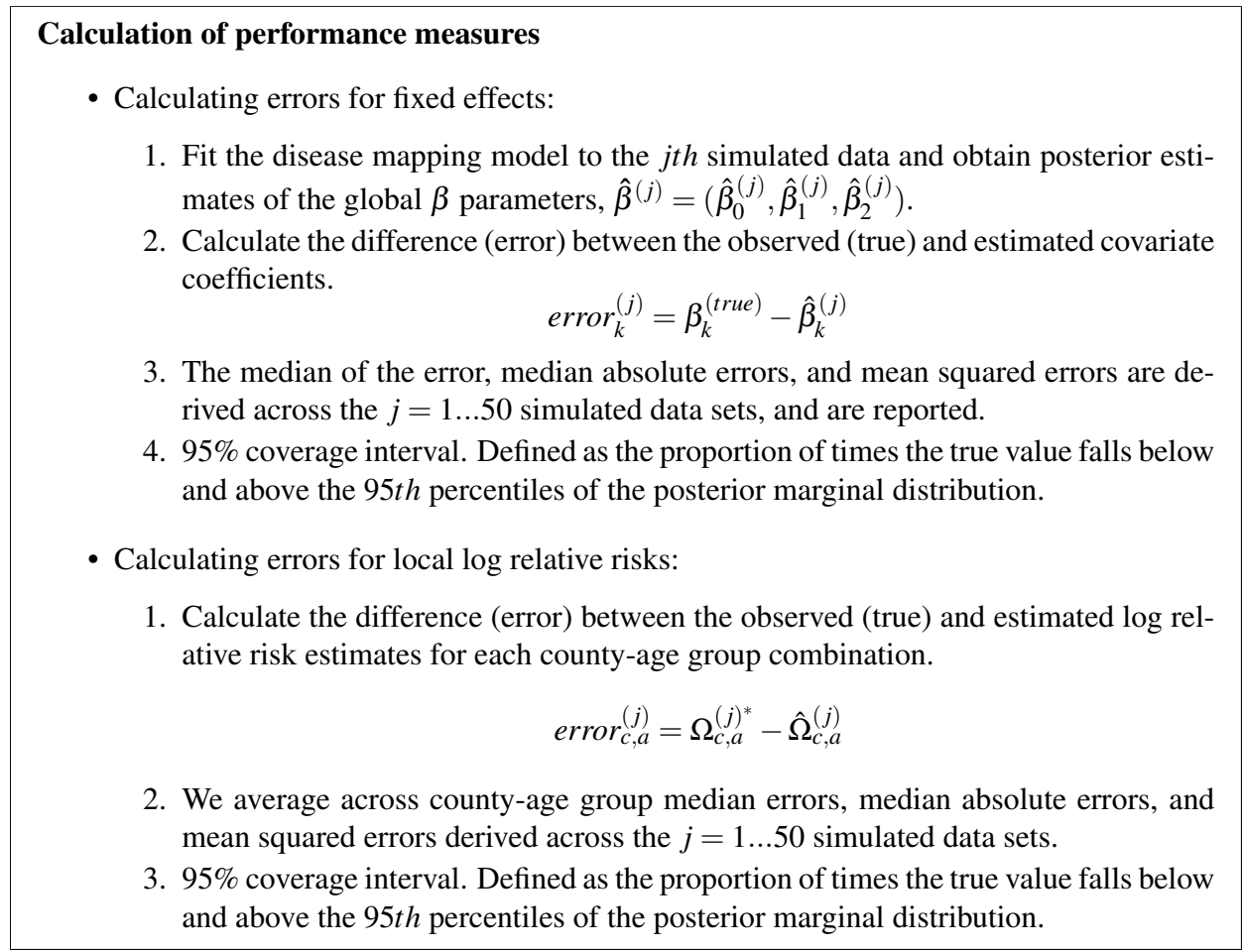

\begin{mdframed}
\textbf{Calculation of performance measures}
\begin{itemize}
\item Calculating errors for fixed effects:
\begin{enumerate}
\item Fit the disease mapping model to the $jth$ simulated data and obtain posterior estimates of the global $\beta$ parameters, $\mathbf{\hat{\beta}}^{(j)} = (\hat{\beta}^{(j)}_{0}, \hat{\beta}^{(j)}_{1}, \hat{\beta}^{(j)}_{2})$.
\item Calculate the difference (error) between the observed (true) and estimated covariate coefficients.
\[error_k^{(j)} = \beta_k^{(true)} - \hat{\beta}_k^{(j)}\]
\item The median of the error, median absolute errors, and mean squared errors are derived across the $j = 1...50$ simulated data sets, and are reported.
\item $95\%$ coverage interval. Defined as the proportion of times the true value falls below and above the $95th$ percentiles of the posterior marginal distribution.
\end{enumerate}

\item Calculating errors for local log relative risks:
\begin{enumerate}
\item Calculate the difference (error) between the observed (true) and estimated log relative risk estimates for each county-age group combination. 
\[error_{c,a}^{(j)} = \Omega_{c,a}^{(j)^*} - \hat{\Omega}_{c,a}^{(j)}\]
\item We average across county-age group median errors, median absolute errors, and mean squared errors derived across the $j = 1...50$ simulated data sets.
\item $95\%$ coverage interval. Defined as the proportion of times the true value falls below and above the $95th$ percentiles of the posterior marginal distribution.
\end{enumerate}
\end{itemize}

\end{mdframed}
\caption{Overview of calculation of error and coverage of prediction intervals in simulation exercise.}
\label{fig:box2}
\end{figure}

\noindent
 From each simulation run, we obtain final estimates of the global $\bm{\hat{\beta}}$ parameters and county-age-group specific log-relative risk estimates $\hat{\Omega}_{c,a}$. To assess model performance, we summarize errors, defined as the difference between the true value and the corresponding model estimate. Outcome measures consisted of median error (ME), median absolute error (MAE), mean squared error (MSE), and $95\%$ coverage intervals. The procedure for calculating summary measures of error is given in Figure \ref{fig:box2}.

\subsection{Simulation results}\label{sec:simres}
 
\subsubsection{Summary measures for global covariate parameters}
We assess bias of posterior estimates using simulation studies described above. In Table \ref{tab:sims1} we present simulation results for the summary measures of errors (bias) related to covariate coefficient estimates comparing naive and BSBE approaches. Overall, the naive and BSBE approaches produce errors close to 0. For the global intercept $\beta_0$, the lowest median errors (MDEs) and median absolute error (MAEs) are produced using the naive approach with the BSBE-ICAR approach yielding similar errors. The BSBE-ACS approach produces notably higher errors compared to the other two approaches (i.e. naive and BSBE-ICAR). Importantly, PEP-based denominators produces the lowest MAE of 0.0153 among the data sources versus the highest MAE 0.0339 associated with the WP denominator, which indicates that PEP based denominators produced lower errors despite being at a disadvantage of having simulated data generated from ACS population counts. Additionally, the global intercept showed lower coverage compared to the expected 95\% coverage intervals indicating overly narrow coverage intervals across all models. For the proportion Black population related coefficient $\beta_1$, the BSBE-ACS model yields a notably lower MAE of 0.0048 compared to 0.0105 from the naive model when using ACS-based denominators, while the naive and BSBE approaches perform similarly for both PEP and WP-based denominators. Lastly, for the ICE related coefficient $\beta_2$, the BSBE-ICAR approach yields a remarkably lower MAE of 0.0065 using PEP denominators compared to 0.0071 from the naive approach. Additionally, the BSBE-ACS approach also shows a lower MAE of 0.0037 compared to the MAE of 0.0081 from the naive model. In brief, the three approaches perform similarly well across the different data sources with negligible differences across the three coefficients.   
%  \begin{figure}[H]
%  \centering
% \includegraphics[width=0.6\textwidth]{fig/fixed_simulation_plots.pdf}
% \caption{True relative risks (black dots), RR estimates and 95\% CIs broken down by BSBE with fixed ACS errors (red), BSBE with an ICAR prior on unknown errors (orange), and Naive (blue) methods, and data source (PEP, ACS, and WP) for selected counties (DeKalb and Taliaferro).}
% \label{fig:simfix}
% \end{figure}

\subsubsection{Summary measures for relative risk estimates}
\noindent 
Table \ref{tab:sims2} gives the summary errors of log-relative risk estimates averaging across all county-age-group combinations. Using PEP and WP denominators, the BSBE-ICAR approach resulted in lower MAEs (0.0205 and 0.0504) compared to the naive approach (0.0209 and 0.0522), respectively.  Using the ACS denominator, lower bias was found using the naive approach. Coverage intervals are similar between naive and BSBE models and were closer to the expected 95\% coverage. These results highlight the need to account for uncertainty associated with data source specific population-at-risk (offset) values, which yielded smaller errors for PEP and WP-based denominators.
\bigskip

\begin{figure}[H]
 \centering
\includegraphics[width=0.8\textwidth]{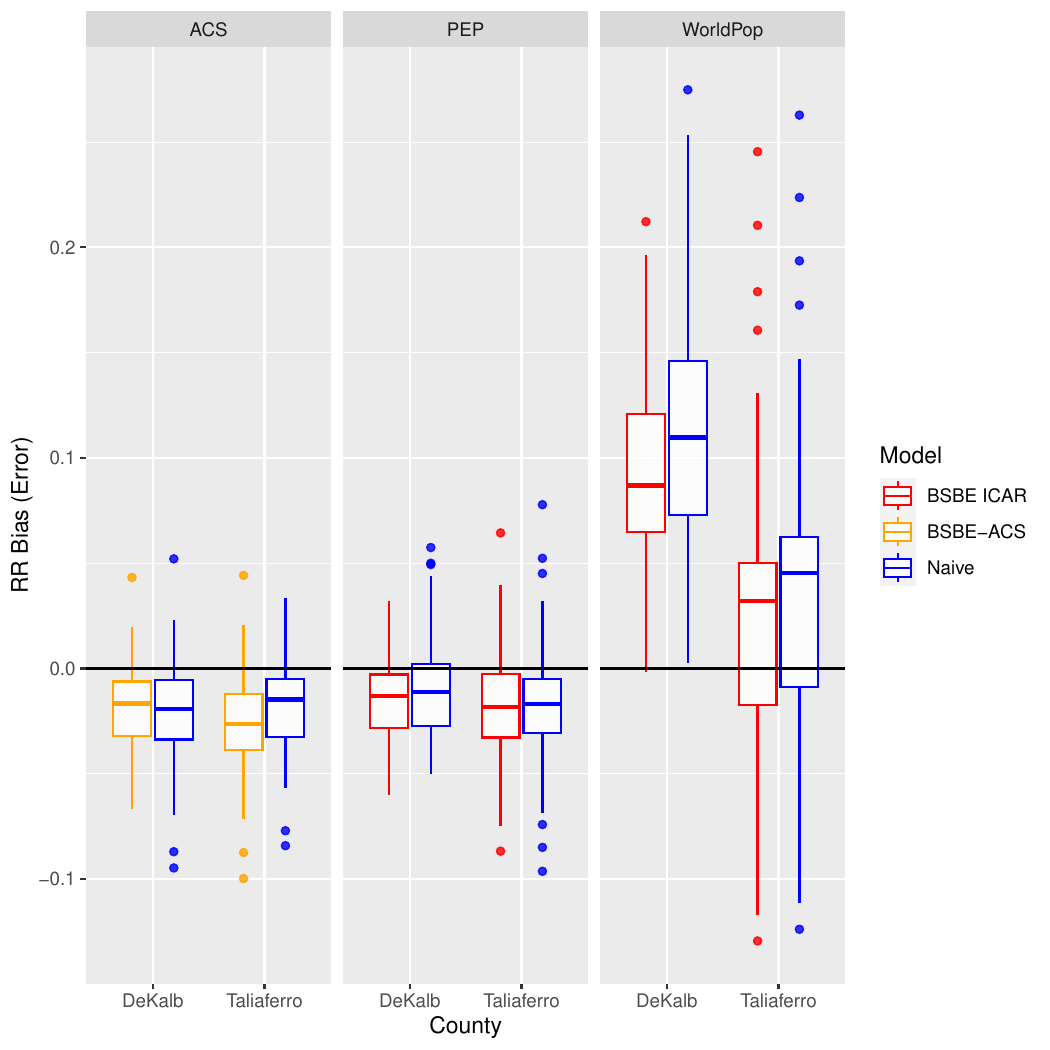}
\caption{Log RR error distributions across all iterations broken down by data source (ACS, PEP, WorldPop) and population-at-risk model (Naive, BSBE-ACS, BSBE-ICAR) for selected counties (DeKalb and Taliaferro).}
\label{fig:simseg}
\end{figure}

\noindent
Figure \ref{fig:simseg} illustrates differences in posterior log RR bias averaged across all iterations comparing BSBE and Naive approaches. We compare trends between two selected counties in Georgia to illustrate differences in results between the larger (DeKalb) and the smaller (Taliaferro) county results. In the case of ACS-based denominators (left column), the Naive and BSBE-ACS approaches produce similar distribution of log RR errors, in which both suffer from a slight bias. The bias is slightly increased using the BSBE-AACS approach. In the case of PEP-based denominators (middle column), the BSBE-ICAR and naive approaches produce similar degrees of bias across RR estimates. Lastly, in the case of WP-based denominators (right column) the BSBE-ICAR approach produces notably lower bias in RR estimates across all iterations for both DeKalb and Taliaferro counties. However, DeKalb county suffers from higher levels of absolute bias compared to the smaller Taliaferro county.  An important summary of these findings is that accounting for population-at-risk related errors does not greatly impact estimates of the covariate coefficients, but is consequential in the estimation of smoothed relative risk estimates. Notably, the method by which population-at-risk error is incorporated also results in differences in smoothed estimates. 

\clearpage\begin{sidewaystable}
\centering
\resizebox{\textwidth}{!}{
\begin{tabular}{|ll|ccccccc|ccccccc|ccccccc|}
  \hline
 & & \multicolumn{7}{c|}{Naive Model} &  \multicolumn{7}{c|}{BSBE-ACS} & \multicolumn{7}{c|}{BSBE-ICAR Model}   \\
  \hline
  \hline
Source & Parameter & ME & MDE & MAE & MSE & LC & UC & IC &ME & MDE & MAE & MSE & LC & UC & IC&  ME & MDE & MAE & MSE & LC & UC & IC  \\ 
 \hline
  \hline
PEP & $\beta_0$ & -0.0179 & -0.0153 & \textbf{0.0153} & 0.0004 & 0 & 0.3 & 0.7 & - &-  &- &- & - & - &- & -0.0187 & -0.0161 & 0.0161 & 0.0005 & 0 & 0.3 & 0.7  \\ 
 ACS & $\beta_0$ & -0.0183 & -0.0181 & \textbf{0.0181} & 0.0004 & 0& 0.3 & 0.7 & -0.0252 & -0.0237 & 0.0237 & 0.0007 & 0 & 0.5 & 0.5 &- &-  &- &- & - & - &-  \\ 
WorldPop & $\beta_0$ & -0.0327 & -0.0339 & 0.0339 & 0.0012 & 0 & 0.8 & 0.2  & - &-  &- &- & - & - &- &-0.0355 & -0.0377 & \textbf{0.0377} & 0.0013 & 0 & 0.9 & 0.1  \\ 
\hline
PEP & $\beta_1$ & 0.0017 & 0.0055 & \textbf{0.0071} & 0.0002 & 0.1 & 0.1 & 0.8  & - &-  &- &- & - & - &-&  0.0016 & 0.0052 & 0.0080 & 0.0002 & 0.1 & 0.1 & 0.8\\ 
ACS & $\beta_1$ & -0.0016 & -0.0014 & 0.0105 & 0.0001 & 0 & 0 & 1 & 0.0005 & 0.0032 &\textbf{0.0048} & 0.0002 & 0.1 & 0.1& 0.8 & - &-  &- &- & - & - &- \\ 
 WorldPop & $\beta_1$ & 0.0021 & 0.0061 & \textbf{0.0079} & 0.0002 & 0.1 & 0.1 & 0.8 & - &-  &- &- & - & - &- & 0.0018 & 0.0057 & 0.0086 & 0.0002 & 0.1 & 0 & 0.9  \\ 
\hline
PEP & $\beta_2$ & 0.0022 & -0.0043 & 0.0071 & 0.0002 & 0.1 & 0 & 0.9  & - &-  &- &- & - & - &- & 0.0021 & -0.0041 & \textbf{0.0065} & 0.0002 & 0.1 & 0 & 0.9 \\ 
ACS & $\beta_2$ & 0.0010 & 0.0005 & 0.0081 & 0.0002 & 0.1 & 0.1 & 0.8 & 0.0033 & -0.0012 & \textbf{0.0077} & 0.0002 & 0.1 & 0 & 0.9 &  - &-  &- &- & - & - &- \\ 
WorldPop & $\beta_2$& 0.0016 & -0.0044 & 0.0064 & 0.0002 & 0.1 & 0 & 0.9 & - &-  &- &- & - & - &-& 0.0013 & -0.0046 & 0.0066 & 0.0002 & 0.1 & 0 & 0.9 \\ 
   \hline
   \hline
\end{tabular}}
\caption{Simulation results for global covariate coefficients. Summary measures include mean error (ME), median error (MDE), median absolute error (MAE), standard deviation of error (SDE), mean square error (MSE), proportion below $2.5\%$ coverage (CL), proportion above $97.5\%$ coverage (UC), and proportion inside $95\%$ coverage (IC). MAE values denoted with bold text denote the lowest MAE in a given row.}
\label{tab:sims1}
\bigskip

\resizebox{\textwidth}{!}{
\begin{tabular}{|l|ccccccc||ccccccc|ccccccc|}
  \hline
 &  \multicolumn{7}{c|}{Naive Model} &  \multicolumn{7}{c|}{BSBE ACS} & \multicolumn{7}{c|}{BSBE-ICAR Model} \\
  \hline
  \hline
Source &  ME & MDE & MAE & MSE & LC & UC & IC &ME & MDE & MAE & MSE & LC & UC & IC&  ME & MDE & MAE & MSE & LC & UC & IC\\ 

\hline
  \hline
PEP & -0.0152 & -0.0156 & 0.0209 & 0.0011 & 0.0003 & 0.0009 & 0.9988& - &-  &- &- & - & - &- & -0.0169 & -0.0169 & 0.0205 & 0.0010 & 0.0001 & 0.0011 & 0.9988  \\ 
 ACS & -0.0171 & -0.0158 & \textbf{0.0187} & 0.0009 & 0.0000 & 0.0072 & 0.9928 & -0.0235 & -0.0236 & 0.0254 & 0.0012 & 0.0001 & 0.0072 & 0.9927 & -& - &-  &- &- & - & -   \\ 
 WorldPop & -0.0224 & -0.0238 & 0.0522 & 0.0075 & 0.0055 & 0.0219 & 0.9725 & - &-  &- &- & - & - &- & -0.0271 & -0.0282 & 0.0504 & 0.0067 & 0.0019 & 0.0232 & 0.9749  \\ 
   \hline
   \hline
\end{tabular}}
\caption{Simulation results for county-age specific relative risk estimates. Summary measures include mean error (ME), median error (MDE), median absolute error (MAE), standard deviation of error (SDE), mean square error (MSE), proportion below $2.5\%$ coverage (CL), proportion above $97.5\%$ coverage (UC), and proportion inside $95\%$ coverage (IC). MAE values denoted with bold text denote the lowest MAE in a given row.}
\label{tab:sims2}
\end{sidewaystable}
 \clearpage

% \subsubsection{Sensitivity Analyses}
% Sensitivity analyses were performed to compare our model assumption of a spatial correlation error structures to a simpler assumption of independence. In the simple hierarchical set-up, we model the unknown offset associated error using a non-informative truncated normal prior denoted $N_{[0,]}()$, and compare results between the BSBE and simple model approach to assess robustness of our model assumptions. We present results of the sensitivity analysis in Appendix Section XX. 

\section{Application of BSBE to estimate small area risk of opioid-related mortality}\label{sec:opapp}
\noindent
Opioid-related mortality is an ongiong public health crisis confronting the 21st century in the United States. Opioid-related drug overdose mortality increased 4-fold between 1999 and 2017 \citep{abdalla}. Previous studies have assessed small area opioid-related mortality rates accounting for variability in space and time characteristics and environments of people who use opioids \citep{sumetsky, klinespat, klinespat2, kline, rossen}. Additionally, the burden of opioid-related mortality is not equally shared across socio-demographic, economic, and geographic characteristics \citep{klinespat, klinespat2}.  A limitation of the previous approaches is that they do not account for the uncertainty associated with local estimates of population size used to derive opioid-related mortality rates at local, state, and national levels. Resulting opioid mortality estimates and associated uncertainties may be inaccurate. We illustrate the use of the BSBE approach to obtain age-group stratified opioid-related mortality rates accounting for population-at-risk uncertainty for 159 counties in Georgia in 2020. We acquired county-level opioid-related deaths stratified by five year age-groups for the state of GA through the Georgia Department of Health (GADPH) \citep{gadph}, and use PEP, ACS, and WP population projections as the denominators. We make comparisons of relative risk estimates across the different denominator data sources also comparing between the BSBE and naive approaches. To inform county-age-group specific estimates of opioid mortality risk, we use covariates that capture racial and economic inequities across age groups and within counties. Specifically, we include the Index of Concentration at Extremes $ICE$ measure which captures economic polarization between White and Black residents within an area \citep{krieger}, further described in \ref{sec:ice}, and the ACS reported proportion Black population out of total population denoted $PropBlack$. We graphically assessed the relationship between these informative covariates and found there is a notable negative relationship in which smaller Black proportions of the county populations were associated with higher ICE measures, which can be seen in \ref{sec:covs}.
Our BSBE model for opioid-related mortality risk is given by Eq. \ref{eq:full} where $y_{c,a}$ denotes the total number of opioid deaths for each $[c,a]$ combination. The hyper-parameter, fixed, and spatially structured and unstructured random effects are hierarchically modeled as defined in Eq. \ref{eq:proc}. We use the Berkson error approach defined in Eqs \ref{eq:pop_pep} - \ref{eq:pop_wp} to model the unknown true population size, and the associated population-at-risk uncertainty $\sigma_{c,a}$, across the denominator types. We compare $[c,a]$ opioid-related mortality risks obtained from the BSBE model to those obtained from a naive model, which assumes the population-at-risk to be fixed, e.g., $log(\gamma_{c,a}) = log\left(n^{(PEP)}_{c,a}\right)$ to illustrate the discrepancy in resulting small area opioid-mortality log-relative risk estimates between the two approaches.
\begin{align}\label{eq:full}
    y_{c,a}|\omega_{c,a} &\sim Poisson(exp(\omega_{c,a}))\\
    \omega_{c,a}& = \mu_{c,a} + log(R) + log(\gamma_{c,a})\nonumber\\
    \mu_{c,a}& = \beta_0 + \beta_1 \cdot PropBlack_{c,a} + \beta_2 \cdot  ICE_{c,a} + \nonumber \\
    & (\sqrt{\rho}\cdot\theta_c^* + \sqrt{1-\rho}\cdot\phi_c^*)\cdot \delta \nonumber
\end{align}

 \subsection{Opioid-related mortality risk in Georgia}\label{sec: res}
 \subsubsection{Global Parameter Estimates}
Table \ref{tab:global} gives posterior mean and median estimates, and 95\% CIs for global hyperparameters comparing naive, BSBE-ACS, and BSBE-ICAR approaches across data sources \footnote{Using PEP and WP denominator data, the BSBE-ACS approach is not applicable}. Using PEP denominators, the global intercept $\beta_0$ was found to be -0.56 (-0.88, -0.29) and -0.58 (-0.90, -0.28) for the naive and BSBE-ICAR, respectively. The covariate coefficient for Proportion Black population $\beta_1$ shows a slight decrease in median estimates from 0.26 (-0.34, 0.92) in the naive approach to 0.20 (-0.45, 0.98) in the BSBE-ICAR approach. The covariate coefficient for the ICE measure $\beta_2$ shows a decrease from 1.58 (0.66, 2.60) in the naive model to 1.47 (2.62) in the BSBE-ICAR model, respectively.  The global spatial correlation parameter $\rho$  shows an increase between the naive to the BSBE-ICAR approach, i.e., 0.02 (0.00, 0.15) to 0.40 (0.02, 0.86), respectively. These results suggest that incorporation of population-at-risk related errors improves distinction between the spatially structure and unstructured terms defined in the underlying process model. Using ACS denominators, there is little difference in global parameter estimates and associated $95\%$ CIs between the Naive and BSBE-ACS approaches also confirming the results found in the simulation study. Additionally, ACS results mimic those from PEP naive and BSBE-ICAR results with exception of the spatial correlation term, which is underestimated at 0.03 (0.00, 0.13) using the BSBE-ACS approach. Lastly, using WP denominators, the covariate coefficient for the Proportion Black population $\beta_1$ has a stark increase from 0.35 (-0.28, 0.99) to 1.03 (0.30, 1.79) comparing naive and BSBE approaches, respectively. The covariate coefficient for the ICE measure $\beta_2$ also shows a drastic increase comparing naive and BSBE approaches, i.e., 1.45 (0.49, 2.45) versus 2.50 (1.42, 3.57). The spatial mixing term is estimated similarly to that of PEP-based denominators. The data-source specific variation $\sigma_{(WP)}$ was found to be 0.2 (0.02, 0.42) indicating that WP suffers from a notable data-source specific error associated with WP estimates.

\begin{table}[ht]
\centering
\resizebox{0.95\textwidth}{!}{
\begin{tabular}{|l|l|rrrrr|rrrrr|rrrrr|}
  \hline
  & &\multicolumn{5}{|c|}{Naive} & \multicolumn{5}{|c|}{BSBE-ACS}  &\multicolumn{5}{|c|}{BSBE-ICAR Model}  \\
  \hline
  \hline
    Source & Parameter & Mean & SD & 2.5\% Q & Median & 97.5\% Q & Mean & SD & 2.5\% Q & Median & 97.5\% Q &Mean & SD & 2.5\% Q & Median & 97.5\% Q\\
  \hline
  \multirow{5}{*}{PEP} & $\beta_0$ & -0.57 & 0.15 & -0.88 & -0.56 & -0.29 &-&-&-&-&-& -0.58 & 0.16 & -0.90 & -0.58 & -0.28  \\ 
&  $\beta_1$ &  0.27 & 0.33 & -0.34 & 0.26 & 0.92 &-&-&-&-&-& 0.21 & 0.36 & -0.45 & 0.20 & 0.97\\ 
& $\beta_2$ & 1.59 & 0.50 & 0.66 & 1.58 & 2.60 &-&-&-&-&-& 1.49 & 0.55 & 0.46 & 1.47 & 2.62  \\ 
 
& $\delta$ & 0.99 & 0.23 & 0.70 & 0.92 & 1.54 & -&-&-&-&-&0.64 & 0.33 & 0.04 & 0.65 & 1.37\\ 
& $\rho$  & 0.04 & 0.04 & 0.00 & 0.02 & 0.15 &-&-&-&-&-& 0.40 & 0.22 & 0.02 & 0.40 & 0.86\\ 
  % $\omega$ & - & - & -&-  & - & 80 & 103.56 & 0.006 &35.3  &370.88  & -& - & - & - & - \\ 
   \hline
     \multirow{5}{*}{ACS} & $\beta_0$ & -0.58 & 0.16 & -0.91 & -0.58 & -0.29 & -0.59 & 0.15 & -0.92 & -0.58 & -0.31 &- & - & - & - & -  \\ 
&  $\beta_1$ & 0.28 & 0.33 & -0.33 & 0.27 & 0.98 & 0.28 & 0.33 & -0.33 & 0.27 & 1.00& - & - & - & - & -  \\ 
& $\beta_2$ & 1.68 & 0.51 & 0.71 & 1.66 & 2.75 & 1.70 & 0.51 & 0.79 & 1.68 & 2.79 &- & - & - & - & - \\ 
& $\delta$ &1.00 & 0.29 & 0.69 & 0.92 & 1.88 & 1.02 & 0.27 & 0.69 & 0.95 & 1.79 &-&-&-&-&- \\ 
& $\rho$  & 0.03 & 0.04 & 0.00 & 0.02 & 0.15 & 0.03 & 0.03 & 0.00 & 0.02 & 0.13&-&-&-&-&-\\ 
\hline
  \multirow{5}{*}{WorldPop} & $\beta_0$ & -0.60 & 0.15 & -0.89 & -0.60 & -0.30 & - & - &-&-&- & -1.03 & 0.18 & -1.38 & -1.02 & -0.70\\
&  $\beta_1$ & 0.35 & 0.32 & -0.28 & 0.35 & 0.99 & -&-&-&-&-& 1.03 & 0.38 & 0.30 & 1.03 & 1.79 \\ 
& $\beta_2$ & 1.45 & 0.50 & 0.49 & 1.45 & 2.45 & -&-&-&-&-&2.49 & 0.55 & 1.42 & 2.50 & 3.57 \\ 
& $\delta$ & 1.07 & 0.30 & 0.71 & 0.99 & 1.87 &-&-&-&-&-& 0.19 & 0.16 & 0.01 & 0.14 & 0.59\\ 
& $\rho$  & 0.03 & 0.04 & 0.00 & 0.02 & 0.16& -&-&-&-&-& 0.42 & 0.29 & 0.01 & 0.38 & 0.96  \\ 
& $\sigma_{(WP)}$ & -&-&-&-&-&-&-&-&-&-& 0.21 & 0.10 & 0.02 & 0.22 & 0.42 \\
\hline
\end{tabular}}
\caption{Global model parameter posterior mean, median, and error estimates and 95\% credible intervals compared between BSBE and Naive model results across data sources: PEP, ACS, and WP.}
\label{tab:global}
\end{table}

 \subsubsection{County level relative risk estimates of opioid-related mortality}
\begin{figure}[H]
\includegraphics[width=1.1\textwidth]{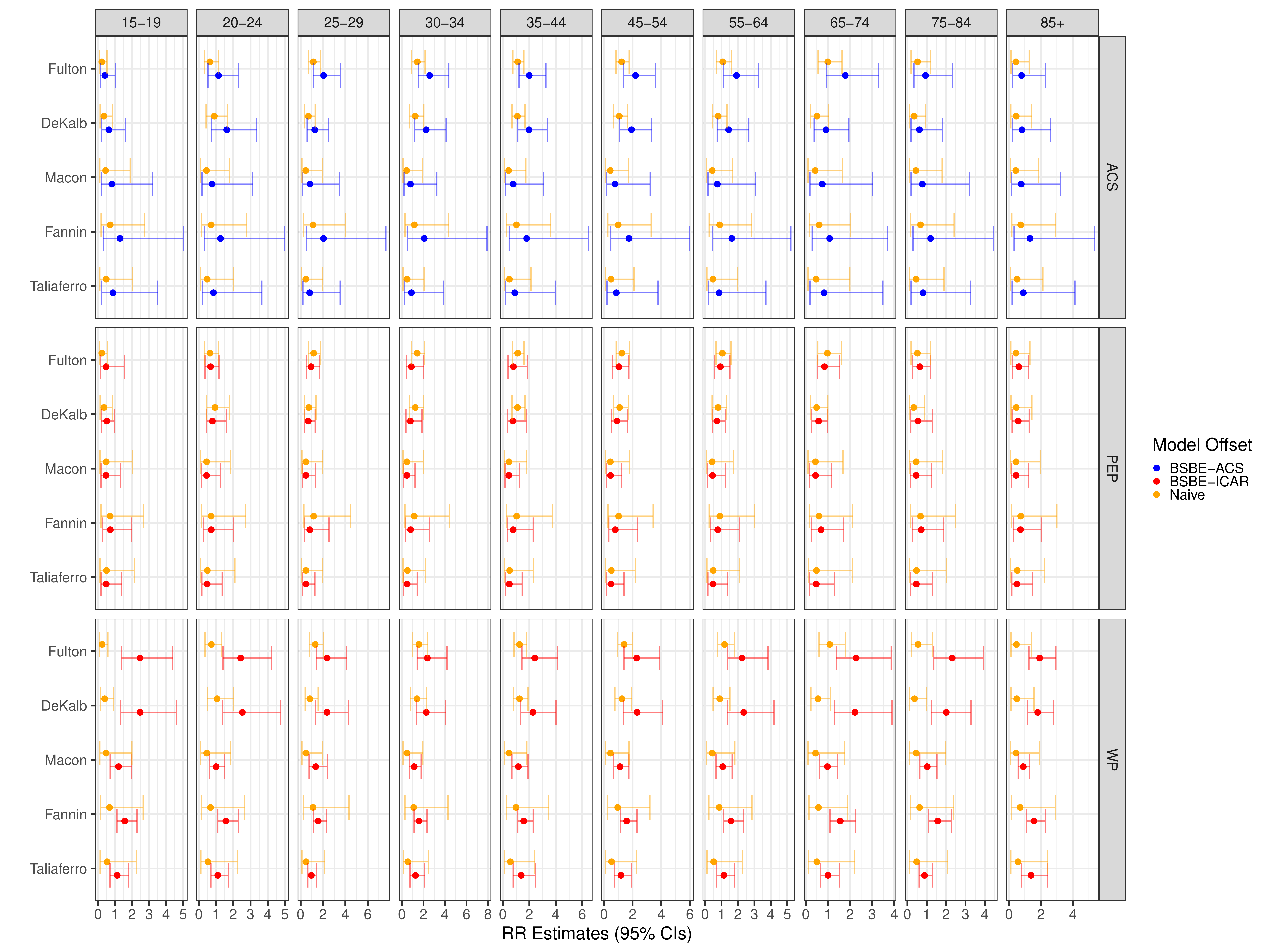}
\caption{Posterior age-group specific median opioid-mortality RR estimates and 95\% credible intervals for selected counties in GA. comparing Naive and BSBE models. Counties are ordered in descending order of population-size.}
\label{fig:selectcos}
\end{figure}

\noindent
Figure \ref{fig:selectcos} shows posterior median opioid-related mortality risk estimates and associated 95\% CIs for selected counties by age-group comparing naive (orange) to BSBE-ACS (blue), and BSBE-ICAR (red) model results. These results confirm the findings illustrated in Figure \ref{fig:simseg} in which the BSBE approaches produces greatly different estimates compared to the naive approach. The great differences in opioid mortality risks across the selected counties suggest that using the naive approach may underestimate local estimates of risk and their associated uncertainties. Additionally, the counties are ordered by total population size from largest (top) to smallest (bottom) to highlight that the larger counties suffer from higher degrees of error using both BSBE approaches in contrast to smaller counties which have higher ACS related errors showing that accounting for spatial variation aids in reducing posterior uncertainty in smaller counties. For example, results for Fulton County in the 35-44 age group based on using ACS denominators, the BSBE-ACS approach produces a higher opioid mortality RR of 2.02 (1.54, 4.03) compared to the naive approach 1.07 (0.80, 1.60). Similarly, when using PEP denominators RR estimates  0.80 (0.40, 1.84) and 1.11 (0.74, 1.61) result from BSBE-ICAR and naive approaches, respectively.  Important to note is that dependent on the method of incorporation of uncertainty, the direction of the RR estimates may change, e.g., RR greater than 1 versus less than 1.  Additionally, the BSBE approaches produce similar results when WP denominators were used which resulted in higher opiod mortality RR estimates and uncertainties compared to the naive approach illustrating the increase in posterior uncertainty when we account for all sources of uncertainty within the disease mapping approach. In contrast, Taliaferro county shows an opioid mortality RR  of 1.01 (0.002, 3.98) using the BSBE-ACS approach versus 0.20 (0.01, 2.03) using the naive approach. When PEP denominators are used, there is a decrease in posterior uncertainty using the BSBE approach, i.e, 0.49 (0.18, 1.46) to 0.51 (0.11, 2.29) using the naive method. A critical improvement is a reduction of uncertainty surrounding RR estimates for smaller counties using the BSBE-ICAR approach compared to the naive approach illustrating the borrowing of information in cases of small sample sizes.

\begin{figure}[H]
\centering
\includegraphics[width=0.95\textwidth]{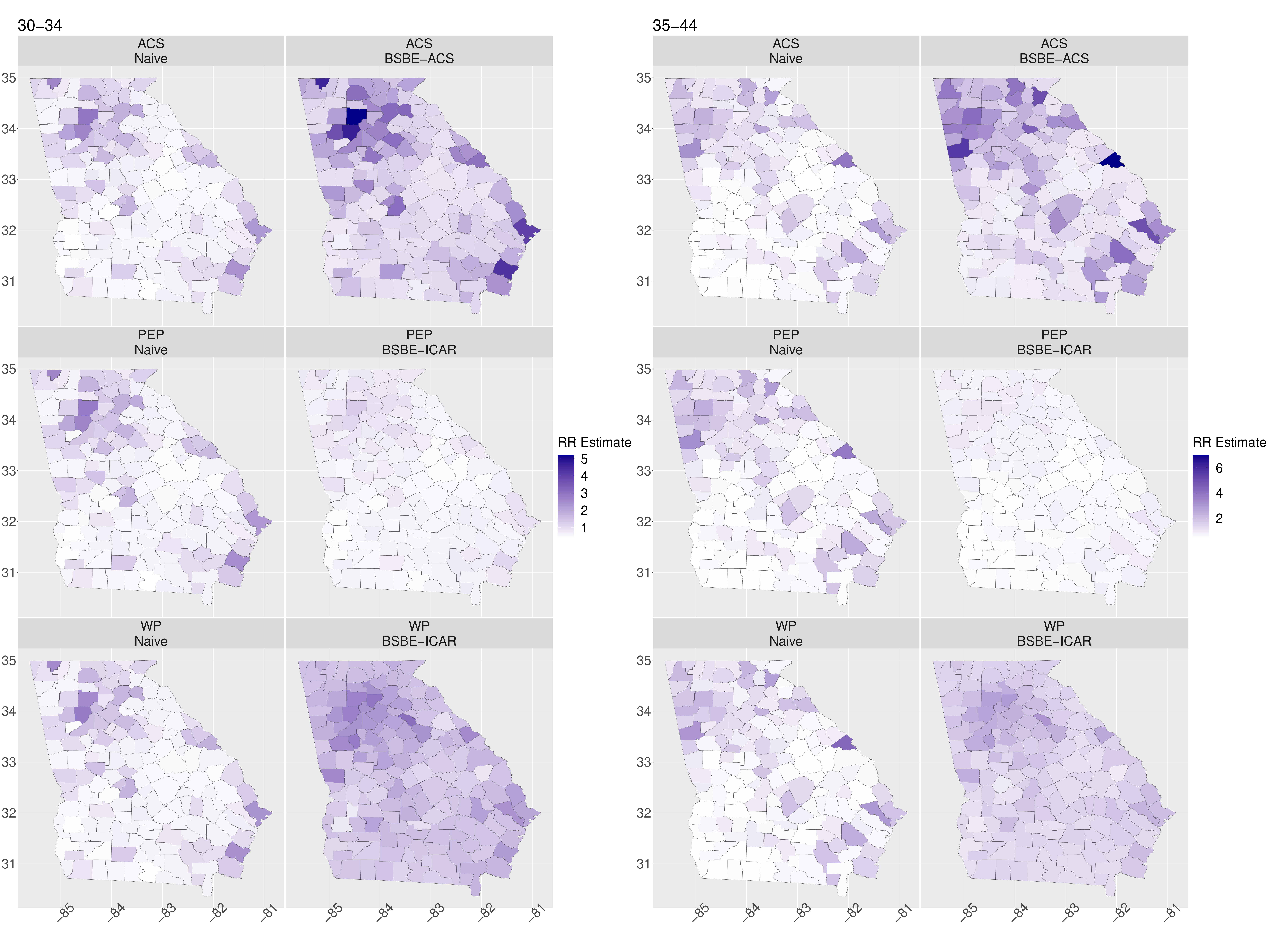}
\caption{Spatial age-group stratified distributions of county-level 2020 opioid-related relative risks (RRs) comparing BSBE and Naive models for Georgia, USA.}
\label{fig:sirresapp}
\end{figure}

 Figure \ref{fig:sirresapp} maps the age-stratified opioid-mortality risks (RRs) for 159 counties in GA in 2020 (for selected age groups)  comparing estimates between model approaches and data sources. The complete set of age-stratified maps can be found in \ref{sec:all} Overall, the BSBE-ICAR approach yields more stable RRs compared to the BSBE-ACS and naive models, with higher risks overall associated with WP denominators. Taking PEP as an example, the naive approach reports the largest opioid-related RR of 3.86 (2.29, 6.15) to be in Richmond County for age-group 35-44. In contrast, the BSBE-ICAR approach reported an RR of 0.76 (0.34, 2.70) for the corresponding county-age-group. The BSBE-ICAR and approach estimates the highest RR estimates of 1.52 to be associated with Columbia County for the 20-24 age-group, respectively. In contrast, the naive model reports the smallest RR of 0.21 (0.07, 0.54) is associated with Fulton County for age-group 15-19, which is increased to 0.47 (0.15, 1.53) by the BSBE-ICAR model. This illustrates the general pattern across PEP and WP in which the BSBE approaches imposes additional smoothing of extremely high or low relative risk estimates due to the incorporation of offset uncertainty. In contrast, the BSBE-ACS approach imposes additional variability compared to the naive approach. Additionally, due to the additional smoothing term in the WP hierarchical model, there is an overall smoothing of RR estimates to age-group specific levels resulting in overall higher relative risk estimates.
 
\section{Discussion}\label{sec:discussion}

Monitoring of small-area geographical population trends in disease and mortality has large scale implications in informing public health policy. The standard approach to obtaining small area disease/mortality estimates may result in biased estimates by ignoring uncertainty associated with population-at-risk estimates. In this paper, we present a Bayesian Spatial Berkson Error approach to incorporate uncertainty associated with population-at-risk (offset) estimates within a disease mapping model across multiple denominator data sources. Incorporation of offset uncertainty is challenging because of the lack of information available on  errors surrounding reported population-at-risk estimates, and the stark differences across data source specific methodologies used to derive small area population count estimates. However, it is necessary to incorporate offset uncertainty (accounting for data source specific methods) in order to obtain accurate small area estimates of risk and associated uncertainty bounds. We assessed incorporation of offset uncertainty using a Berkson error model, and provided data-source specific mechanisms to incorporate that uncertainty across multiple data sources (PEP, ACS, and WP). Importantly, we motivate our model choice for unknown offset errors based on the spatial distributions of ACS reported standard errors. Methodologies that aim to identify small areas and populations of greater need to develop targeted interventions should consider these challenges within the methodological framework. \\

Model and simulation results suggest that the naive and BSBE-ICAR approaches produce similar estimates of risk when assumptions of normality have been met. Additionally, 95\% uncertainty bounds surrounding posterior RR estimates are larger using the naive approach for smaller counties, which benefit from sharing of information through hierarchical structures on error terms. These findings indicate an increased smoothing of RR estimates using the BSBE-ICAR approach due to sharing of information across small areas via the hierarchical Berkson error model. This model improves upon the limitations of the naive approach in which we assume denominators to be fixed and known. We illustrated how our proposed model may be used in the assessment of small area risks of opioid-related mortality for 159 counties in GA across 10 age-groups. The BSBE models produces opioid-mortality risks with smaller degrees of posterior uncertainty (i.e., reduction in uncertainty bounds) for smaller counties such as Taliaferro and Fannin, but additionally imposed increased smoothing of risk estimates for counties that suffered from extremely high or low values (i.e., Richmond County) using the naive model approach. More information is needed to understand why some counties and age-groups have extremely high or low risk values, and how that may be informative and guide improvements in opioid-risk estimation. \\

This work makes several contributions. First, we introduce a novel disease mapping approach that can account for uncertainty associated with denominator data across multiple data sources. To our knowledge, no study has addressed this limitation within the disease mapping framework across different data types. Our model approach has been developed to do the following: (1) Describe the different degrees of error across multiple data types, i.e, (USCB data sources) as well as other alternative machine learning based population estimates such as WorldPop, and (2) Extend our model framework to account for varying degrees of error associated with various denominator data. Incorporation of the Berkson error within disease mapping allows for fine tuning, smoothing, and borrowing of information to reflect the strengths and weaknesses of the data products.  Although we applied our approach to opioid-related mortality for 159 counties in Georgia, this approach can be applied to other states within the U.S. and other health indicators to obtain estimates of disease or mortality risks stratified across different socio-demographic populations and varying geographic resolutions, i.e., census tracts. 
\bigskip

 \noindent
\textbf{Acknowledgements:} The authors are very grateful to all the members of the Spatial Uncertainty Research Team, from the Emory Rollins School of Public Health, Harvard T.H. Chan School of Public Health, and the UK Small Area Health Statistics Unit (SAHSU) at Imperial College London's School of Public Health, for the support of this work.
 \bigskip

\noindent
\textbf{Funding:} This work, led by Lance A. Waller, was funded by the US National Institutes for Health (NIH) (\#RO1HD092580). FBP acknowledges support from Health Data Research UK(HDR UK) and the UK National Institute for Health Research (NIHR) Imperial Biomedical Research Centre. FBP is a member of the NIHR Health Protection Research Units in Chemical and Radiation Threats and Hazards, and in Environmental Exposures and Health, which are partnerships between UK Health Safety Agency and Imperial College London funded by the UK NIHR.

 \bibliography{MyLibrary}

\begin{thebibliography}{46}
\expandafter\ifx\csname natexlab\endcsname\relax\def\natexlab#1{#1}\fi
\providecommand{\url}[1]{\texttt{#1}}
\providecommand{\href}[2]{#2}
\providecommand{\path}[1]{#1}
\providecommand{\DOIprefix}{doi:}
\providecommand{\ArXivprefix}{arXiv:}
\providecommand{\URLprefix}{URL: }
\providecommand{\Pubmedprefix}{pmid:}
\providecommand{\doi}[1]{\href{http://dx.doi.org/#1}{\path{#1}}}
\providecommand{\Pubmed}[1]{\href{pmid:#1}{\path{#1}}}
\providecommand{\bibinfo}[2]{#2}
\ifx\xfnm\relax \def\xfnm[#1]{\unskip,\space#1}\fi
%Type = Book
\bibitem[{{Waller, L.A.} and {Gotway, C.A.}(2004)}]{waller_gotway}
\bibinfo{author}{{Waller, L.A.}}, \bibinfo{author}{{Gotway, C.A.}},
  \bibinfo{title}{Applied Spatial Statistics for Public Health Data},
  \bibinfo{publisher}{John Wiley \& Sons, Inc.}, \bibinfo{year}{2004}.
  \DOIprefix\doi{10.1002/0471662682}.
%Type = Article
\bibitem[{{Wakefield, J.}(2007)}]{jon_wakefield_disease_2007}
\bibinfo{author}{{Wakefield, J.}},
\newblock \bibinfo{title}{Disease mapping and spatial regression with count
  data},
\newblock \bibinfo{journal}{Biostatistics} \bibinfo{volume}{8}
  (\bibinfo{year}{2007}) \bibinfo{pages}{158--183}.
  \DOIprefix\doi{10.1093/biostatistics/kx1008}.
%Type = Article
\bibitem[{{United States Census Bureau}(2012)}]{census}
\bibinfo{author}{{United States Census Bureau}},
\newblock \bibinfo{title}{Decennial census: Complete technical documentation}
  (\bibinfo{year}{2012}). \URLprefix
  \url{https://www.census.gov/programs-surveys/decennial-census/technical-documentation/complete-technical-documents.html}.
%Type = Article
\bibitem[{{U.S. Census Bureau}(2018)}]{us_census_bureau_understanding_2018}
\bibinfo{author}{{U.S. Census Bureau}},
\newblock \bibinfo{title}{Understanding and {Using} {American} {Community}
  {Survey} {Data}: {What} all data users need to know}  (\bibinfo{year}{2018}).
  \URLprefix \url{census.gov}.
%Type = Article
\bibitem[{{Population Estimation Program, U.S. Census Bureau}(2019)}]{pep}
\bibinfo{author}{{Population Estimation Program, U.S. Census Bureau}},
\newblock \bibinfo{title}{Methodology for the united states population
  estimates: Vintage 2019}  (\bibinfo{year}{2019}). \URLprefix
  \url{https://www2.census.gov/programs-surveys/popest/technical-documentation/methodology/2010-2019/natstcopr-methv2.pdf},
  \bibinfo{note}{accessed on 07-05-2021}.
%Type = Misc
\bibitem[{WorldPop(2020)}]{worldpop}
\bibinfo{author}{WorldPop}, \bibinfo{title}{Worldpop griddded population
  estimate datasets and tools. how are they different and which should i use?},
  \bibinfo{howpublished}{\url{https://www.worldpop.org/methods/populations}},
  \bibinfo{year}{2020}. \URLprefix
  \url{https://www.worldpop.org/methods/populations.}
%Type = Article
\bibitem[{{Starsinic, M.} and {Albright, K.}(2001)}]{CC}
\bibinfo{author}{{Starsinic, M.}}, \bibinfo{author}{{Albright, K.}},
\newblock \bibinfo{title}{Coverage and completeness in the census 2000
  supplementary survey},
\newblock \bibinfo{journal}{Demographic Statistical Methods Division, U.S.
  Census Bureau, Washington, DC}  (\bibinfo{year}{2001}). \URLprefix
  \url{https://www.census.gov/content/dam/Census/library/working-papers/2002/acs/2002_Starsinic_01.pdf}.
%Type = Misc
\bibitem[{{U.S. Census Bureau}(2004)}]{us2004accuracy}
\bibinfo{author}{{U.S. Census Bureau}}, \bibinfo{title}{Accuracy and coverage
  evaluation of census 2000},
  \bibinfo{howpublished}{\url{https://www2.census.gov/programs-surveys/decennial/2000/technical-documentation/coverage-evaluation/dssd03-dm.pdf}},
  \bibinfo{year}{2004}. \URLprefix
  \url{https://www2.census.gov/programs-surveys/decennial/2000/technical-documentation/coverage-evaluation/dssd03-dm.pdf}.
%Type = Misc
\bibitem[{{U.S. Census Bureau: Measures of Nonsampling Error}(2015)}]{nonsamp}
\bibinfo{author}{{U.S. Census Bureau: Measures of Nonsampling Error}},
  \bibinfo{title}{Statistical quality standard d3: Producing measures and
  indicators of nonsampling error},
  \bibinfo{howpublished}{\url{https://www.census.gov/about/policies/quality/standards/standardd3.html}},
  \bibinfo{year}{2015}. \URLprefix
  \url{https://www.census.gov/about/policies/quality/standards/standardd3.html}.
%Type = Book
\bibitem[{{Preston, S.} et~al.(2000){Preston, S.}, {Heuveline, P.}, and
  {Guillot, M.}}]{demog}
\bibinfo{author}{{Preston, S.}}, \bibinfo{author}{{Heuveline, P.}},
  \bibinfo{author}{{Guillot, M.}}, \bibinfo{title}{Demography; Measuring and
  Modeling Population Processes}, \bibinfo{edition}{first} ed.,
  \bibinfo{publisher}{Wiley Blackwell}, \bibinfo{year}{2000}.
%Type = Article
\bibitem[{{U.S. Census
  Bureau}(2014{\natexlab{a}})}]{us_census_bureau_american_2014-2}
\bibinfo{author}{{U.S. Census Bureau}},
\newblock \bibinfo{title}{American {Community} {Survey}: {Design} and
  {Methodology} {Chapter} 11: {Weighting} and {Estimation}},
\newblock \bibinfo{journal}{U.S. Government Printing Office}
  (\bibinfo{year}{2014}{\natexlab{a}}). \URLprefix
  \url{https://www.census.gov/programs-surveys/acs/methodology/weighting-and-estimation.html}.
%Type = Article
\bibitem[{{U.S. Census
  Bureau}(2014{\natexlab{b}})}]{us_census_bureau_american_2014-1}
\bibinfo{author}{{U.S. Census Bureau}},
\newblock \bibinfo{title}{The {American} {Community} {Survey}. {Design} and
  {Methodology} {Chapter} 4: {Sample Design} and {Selection}},
\newblock \bibinfo{journal}{U.S. Government Printing Office}
  (\bibinfo{year}{2014}{\natexlab{b}}). \URLprefix
  \url{https://www.census.gov/programs-surveys/acs/methodology/design-and-methodology.html}.
%Type = Misc
\bibitem[{{U.S. Census Bureau}(2009)}]{us_census_bureau_understanding_2009}
\bibinfo{author}{{U.S. Census Bureau}}, \bibinfo{title}{A compass for
  understanding and using american community survey data: What researchers need
  to know.},
  \bibinfo{howpublished}{\url{https://www.census.gov/content/dam/Census/library/publications.2009/acs/ACSResearch.pdf}},
  \bibinfo{year}{2009}. \URLprefix
  \url{https://www.census.gov/content/dam/Census/library/publications.2009/acs/ACSResearch.pdf}.
%Type = Article
\bibitem[{{Nethery, R.} et~al.(2021){Nethery, R.}, {Rushovich, T.}, {Peterson,
  E}, {Chen, J.}, {Waterman, P.}, {Krieger, N.}, {Waller, L.}, and {Coull,
  B.}}]{nethery}
\bibinfo{author}{{Nethery, R.}}, \bibinfo{author}{{Rushovich, T.}},
  \bibinfo{author}{{Peterson, E}}, \bibinfo{author}{{Chen, J.}},
  \bibinfo{author}{{Waterman, P.}}, \bibinfo{author}{{Krieger, N.}},
  \bibinfo{author}{{Waller, L.}}, \bibinfo{author}{{Coull, B.}},
\newblock \bibinfo{title}{Comparing the performance of three census tract
  denominator sources for real-time disease incidence modeling: Us decennial
  census, american community survey, and worldpop},
\newblock \bibinfo{journal}{Health and Place}  (\bibinfo{year}{2021}).
%Type = Article
\bibitem[{{Stevens, Forrest R.} et~al.(2015){Stevens, Forrest R.}, {Gaughan,
  Andrea E.}, {Linard, Catherine}, and {Tatem, Andrew J.}}]{stevens}
\bibinfo{author}{{Stevens, Forrest R.}}, \bibinfo{author}{{Gaughan, Andrea
  E.}}, \bibinfo{author}{{Linard, Catherine}}, \bibinfo{author}{{Tatem, Andrew
  J.}},
\newblock \bibinfo{title}{Disaggregating census data for population mapping
  using random forests with remotely-sensed and ancillary data},
\newblock \bibinfo{journal}{PLOS ONE} \bibinfo{volume}{10}
  (\bibinfo{year}{2015}). \URLprefix
  \url{https://journals.plos.org/plosone/article?id=10.1371/journal.pone.0107042}.
%Type = Article
\bibitem[{{Tatem, Andrew J.} et~al.(2013){Tatem, Andrew J.}, {Garcia A.J.},
  {Snow, R.W.}, {Noor, A.M.}, {Gaughan, A.E.}, and {Gilbert, M.}}]{tatem2}
\bibinfo{author}{{Tatem, Andrew J.}}, \bibinfo{author}{{Garcia A.J.}},
  \bibinfo{author}{{Snow, R.W.}}, \bibinfo{author}{{Noor, A.M.}},
  \bibinfo{author}{{Gaughan, A.E.}}, \bibinfo{author}{{Gilbert, M.}},
\newblock \bibinfo{title}{Millenium development health metrics: where do
  africa's children and womn of childbearing age live?},
\newblock \bibinfo{journal}{Population Health Metrics} \bibinfo{volume}{11}
  (\bibinfo{year}{2013}) \bibinfo{pages}{1--11}. \URLprefix
  \url{https://pophealthmetrics.biomedcentral.com/articles/10.1186/1478-7954-11-11}.
%Type = Book
\bibitem[{{Gustafson, P.}(2004)}]{gustafson}
\bibinfo{author}{{Gustafson, P.}}, \bibinfo{title}{Measurement Error and
  Misclassification in Statistics and Epidemiology: Impacts and Bayesian
  Adjustments}, \bibinfo{publisher}{Chapman \& Hall/ CRC},
  \bibinfo{year}{2004}.
%Type = Book
\bibitem[{{Carroll, R.J.} et~al.(2006){Carroll, R.J.}, {Ruppert, D.},
  {Stefanski, L.A.}, and {Crainiceanu,
  C.M.}}]{raymond_j_carroll_measurement_2006}
\bibinfo{author}{{Carroll, R.J.}}, \bibinfo{author}{{Ruppert, D.}},
  \bibinfo{author}{{Stefanski, L.A.}}, \bibinfo{author}{{Crainiceanu, C.M.}},
  \bibinfo{title}{Measurement {Error} in {Nonlinear} {Models}},
  \bibinfo{edition}{second} ed., \bibinfo{publisher}{Chapman \& Hall/CRC},
  \bibinfo{year}{2006}.
%Type = Article
\bibitem[{{Huque, M.H.} et~al.(2016){Huque, M.H.}, {Bondell, H.D.}, {Carroll,
  R.J.}, and {Ryan, L.M.}}]{huque}
\bibinfo{author}{{Huque, M.H.}}, \bibinfo{author}{{Bondell, H.D.}},
  \bibinfo{author}{{Carroll, R.J.}}, \bibinfo{author}{{Ryan, L.M.}},
\newblock \bibinfo{title}{Spatial regression with covariate measurement error:
  A semiparametric approach},
\newblock \bibinfo{journal}{Biometrics} \bibinfo{volume}{72}
  (\bibinfo{year}{2016}) \bibinfo{pages}{678--86}.
%Type = Article
\bibitem[{{Huque, M.H.} et~al.(2014){Huque, M.H.}, {Bondell, H.D.}, and {Ryan,
  L.M.}}]{huque2}
\bibinfo{author}{{Huque, M.H.}}, \bibinfo{author}{{Bondell, H.D.}},
  \bibinfo{author}{{Ryan, L.M.}},
\newblock \bibinfo{title}{On the impact of covariate measurement error on
  spatial regression modelling},
\newblock \bibinfo{journal}{Environmetrics} \bibinfo{volume}{25}
  (\bibinfo{year}{2014}) \bibinfo{pages}{560--70}.
%Type = Article
\bibitem[{{Li, Y.} et~al.(2009){Li, Y.}, {Tang, H.}, and {Lin, X.}}]{li}
\bibinfo{author}{{Li, Y.}}, \bibinfo{author}{{Tang, H.}},
  \bibinfo{author}{{Lin, X.}},
\newblock \bibinfo{title}{Spatial linear mixed models with covariate
  measurement error},
\newblock \bibinfo{journal}{Stastica Sinica} \bibinfo{volume}{19}
  (\bibinfo{year}{2009}) \bibinfo{pages}{1077--93}.
%Type = Article
\bibitem[{{Zhang, K.} et~al.(2021){Zhang, K.}, {Liu, J.}, {Zhang, P.}, and
  {Carroll, R.J.}}]{zhang}
\bibinfo{author}{{Zhang, K.}}, \bibinfo{author}{{Liu, J.}},
  \bibinfo{author}{{Zhang, P.}}, \bibinfo{author}{{Carroll, R.J.}},
\newblock \bibinfo{title}{Bayesian adjustment for measurement error in an
  offset variable in a poisson regression model},
\newblock \bibinfo{journal}{Statistics Modelling}  (\bibinfo{year}{2021})
  \bibinfo{pages}{1--18}.
%Type = Article
\bibitem[{{Josey, K.P.} et~al.(2023){Josey, K.P.}, {deSouza, P.}, {Wu, X.},
  {Braun, D.}, and {Nethery, R.}}]{josey}
\bibinfo{author}{{Josey, K.P.}}, \bibinfo{author}{{deSouza, P.}},
  \bibinfo{author}{{Wu, X.}}, \bibinfo{author}{{Braun, D.}},
  \bibinfo{author}{{Nethery, R.}},
\newblock \bibinfo{title}{Estimating a causal exposure response function with a
  continuous error-prone exposure: A study of fine particulate matter and
  all-cause mortality},
\newblock \bibinfo{journal}{JABES} \bibinfo{volume}{28} (\bibinfo{year}{2023})
  \bibinfo{pages}{20--41}. \URLprefix
  \url{https://doi.org/10.1007/s13253-022-00508-z}.
%Type = Article
\bibitem[{{Peterson, E.N.} et~al.(2023){Peterson, E.N.}, {Nethery, R.C.},
  {Padellini, T.}, {Chen, J.T.}, {Coull, B.A.}, {Piel, F.B.}, {Wakefield, J.},
  {Blangiardo, M.}, and {Waller, L.A.}}]{peterson}
\bibinfo{author}{{Peterson, E.N.}}, \bibinfo{author}{{Nethery, R.C.}},
  \bibinfo{author}{{Padellini, T.}}, \bibinfo{author}{{Chen, J.T.}},
  \bibinfo{author}{{Coull, B.A.}}, \bibinfo{author}{{Piel, F.B.}},
  \bibinfo{author}{{Wakefield, J.}}, \bibinfo{author}{{Blangiardo, M.}},
  \bibinfo{author}{{Waller, L.A.}},
\newblock \bibinfo{title}{A bayesian hierarchical small area population model
  accounting for data source specific methodologies from american community
  survey, population estimates program, and decennial census data},
\newblock \bibinfo{journal}{Annals of Applied Statistics} \bibinfo{volume}{TBD}
  (\bibinfo{year}{2023}) \bibinfo{pages}{TBD}.
%Type = Article
\bibitem[{{Besag, J.} et~al.(1991){Besag, J.}, {York, J.}, and {Mollie,
  A.}}]{bym}
\bibinfo{author}{{Besag, J.}}, \bibinfo{author}{{York, J.}},
  \bibinfo{author}{{Mollie, A.}},
\newblock \bibinfo{title}{Bayesian image restoration with two applications in
  spatial statistics (with discussion)},
\newblock \bibinfo{journal}{Annals of the Institute of Statistical Mathematics}
  \bibinfo{volume}{43} (\bibinfo{year}{1991}) \bibinfo{pages}{1––59}.
%Type = Article
\bibitem[{{Riebler, A.} et~al.(2016){Riebler, A.}, {Sørbye, S.H.}, {Simpson,
  D.}, and {Rue, H.}}]{andrea_riebler_intuitive_2016}
\bibinfo{author}{{Riebler, A.}}, \bibinfo{author}{{Sørbye, S.H.}},
  \bibinfo{author}{{Simpson, D.}}, \bibinfo{author}{{Rue, H.}},
\newblock \bibinfo{title}{An {Intuitive} {Bayesian} {Spatial} {Model} for
  {Disease} {Mapping} {That} {Accounts} for {Scaling}},
\newblock \bibinfo{journal}{Statistical Methods in Medical Research}
  \bibinfo{volume}{25} (\bibinfo{year}{2016}) \bibinfo{pages}{1145--65}.
%Type = Article
\bibitem[{{Zhang, K.} et~al.(2021){Zhang, K.}, {Liu, J.}, {Zhang, P.}, and
  {Carroll, R.J.}}]{kline}
\bibinfo{author}{{Zhang, K.}}, \bibinfo{author}{{Liu, J.}},
  \bibinfo{author}{{Zhang, P.}}, \bibinfo{author}{{Carroll, R.J.}},
\newblock \bibinfo{title}{Estimating the burden of the opioid epidemic for
  adults and adolescents in ohio counties},
\newblock \bibinfo{journal}{Biometrics} \bibinfo{volume}{77}
  (\bibinfo{year}{2021}) \bibinfo{pages}{765--777}.
  \DOIprefix\doi{10.1111/biom.13295}.
%Type = Article
\bibitem[{{Kline, D.} et~al.(2021){Kline, D.}, {Pan, Y.}, and {Hepler,
  S.}}]{klinespat}
\bibinfo{author}{{Kline, D.}}, \bibinfo{author}{{Pan, Y.}},
  \bibinfo{author}{{Hepler, S.}},
\newblock \bibinfo{title}{Spatio-temporal trends in opioid overdose deaths by
  race for counties in ohio},
\newblock \bibinfo{journal}{Epidemiology} \bibinfo{volume}{32}
  (\bibinfo{year}{2021}) \bibinfo{pages}{295--302}.
%Type = Article
\bibitem[{{Hepler, S.} et~al.(2021){Hepler, S.}, {Waller, L.A.}, and {Kline,
  D.M.}}]{klinespat2}
\bibinfo{author}{{Hepler, S.}}, \bibinfo{author}{{Waller, L.A.}},
  \bibinfo{author}{{Kline, D.M.}},
\newblock \bibinfo{title}{A multivariate spatiotemporal change-point model of
  opioid overdose deaths in ohio},
\newblock \bibinfo{journal}{Annals of Applied Statistics} \bibinfo{volume}{15}
  (\bibinfo{year}{2021}) \bibinfo{pages}{1329--1342}.
%Type = Article
\bibitem[{{Rossen, L.M.} et~al.(2014){Rossen, L.M.}, {Khan, D.}, and {Warner,
  M.}}]{rossen}
\bibinfo{author}{{Rossen, L.M.}}, \bibinfo{author}{{Khan, D.}},
  \bibinfo{author}{{Warner, M.}},
\newblock \bibinfo{title}{Hot spots in mortality from drug poisoning in the
  united states, 2007-2009},
\newblock \bibinfo{journal}{Health and Place} \bibinfo{volume}{26}
  (\bibinfo{year}{2014}) \bibinfo{pages}{14--20}. \DOIprefix\doi{10.1016}.
%Type = Article
\bibitem[{{Spielman, S.} and {Folch, D.}(2015)}]{seth_e_spielman_reducing_2015}
\bibinfo{author}{{Spielman, S.}}, \bibinfo{author}{{Folch, D.}},
\newblock \bibinfo{title}{Reducing {Uncertainty} in the {American} {Community}
  {Survey} through {Data}-{Driven} {Regionalization}},
\newblock \bibinfo{journal}{PLos ONE} \bibinfo{volume}{10}
  (\bibinfo{year}{2015}) \bibinfo{pages}{1--21}.
%Type = Article
\bibitem[{{Spielman, S.} et~al.(2014){Spielman, S.}, {Folch, D.}, and {Nagle,
  N.}}]{spielman_s_patterns_2014}
\bibinfo{author}{{Spielman, S.}}, \bibinfo{author}{{Folch, D.}},
  \bibinfo{author}{{Nagle, N.}},
\newblock \bibinfo{title}{Patterns and causes of uncertainty in the {American}
  {Community} {Survey}},
\newblock \bibinfo{journal}{Applied Geography} \bibinfo{volume}{46}
  (\bibinfo{year}{2014}) \bibinfo{pages}{147--157}.
%Type = Article
\bibitem[{{Starsinic, M.} and {Tersine Jr., A.}(2007)}]{starsinic}
\bibinfo{author}{{Starsinic, M.}}, \bibinfo{author}{{Tersine Jr., A.}},
\newblock \bibinfo{title}{Analysis of variance estimates from {American}
  {Community} {Survey} multiyear estimates},
\newblock \bibinfo{journal}{In Proceedings of the section of survey research
  methods. Alexandria, VA: American Statistical Association}
  (\bibinfo{year}{2007}) \bibinfo{pages}{3011--3017}.
%Type = Article
\bibitem[{{Gelman, A.}(2006)}]{andrew_gelman_prior_2006}
\bibinfo{author}{{Gelman, A.}},
\newblock \bibinfo{title}{Prior distributions for variance parameters in
  hierarchical models},
\newblock \bibinfo{journal}{Bayesian Analysis}  (\bibinfo{year}{2006})
  \bibinfo{pages}{515--533}.
%Type = Misc
\bibitem[{{Walker, K.}(2020)}]{walker}
\bibinfo{author}{{Walker, K.}}, \bibinfo{title}{tidycensus: Load us census
  boundary and attribute data as `tidyverse'. r package version 0.9.9.2},
  \bibinfo{howpublished}{\url{https://walker-data.com/tidycensus/articles/basic-usage.html}},
  \bibinfo{year}{2020}. \URLprefix
  \url{https://walker-data.com/tidycensus/articles/basic-usage.html},
  \bibinfo{note}{accessed: 08-10-2022}.
%Type = Article
\bibitem[{{de Valpine} et~al.(2017){de Valpine}, Turek, Paciorek,
  Anderson-Bergman, {Temple Lang}, and Bodik}]{nimble}
\bibinfo{author}{P.~{de Valpine}}, \bibinfo{author}{D.~Turek},
  \bibinfo{author}{C.~Paciorek}, \bibinfo{author}{C.~Anderson-Bergman},
  \bibinfo{author}{D.~{Temple Lang}}, \bibinfo{author}{R.~Bodik},
\newblock \bibinfo{title}{Programming with models: writing statistical
  algorithms for general model structures with {NIMBLE}},
\newblock \bibinfo{journal}{Journal of Computational and Graphical Statistics}
  \bibinfo{volume}{26} (\bibinfo{year}{2017}) \bibinfo{pages}{403--413}.
  \DOIprefix\doi{10.1080/10618600.2016.1172487}.
%Type = Article
\bibitem[{Plummer(2017)}]{plummer_jags_2017}
\bibinfo{author}{M.~Plummer},
\newblock \bibinfo{title}{{JAGS}: {A} program for analysis of {Bayesian}
  graphical models using {Gibbs} sampling}  (\bibinfo{year}{2017}).
%Type = Article
\bibitem[{Gelman and Rubin(1992)}]{gelman_inference_1992}
\bibinfo{author}{A.~Gelman}, \bibinfo{author}{D.~B. Rubin},
\newblock \bibinfo{title}{Inference from {Iterative} {Simulation} {Using}
  {Multiple} {Sequences}},
\newblock \bibinfo{journal}{Statistical Science} \bibinfo{volume}{7}
  (\bibinfo{year}{1992}) \bibinfo{pages}{457--472}. \URLprefix
  \url{http://www.jstor.org/stable/2246093}, \bibinfo{note}{publisher:
  Institute of Mathematical Statistics}.
%Type = Article
\bibitem[{{Vehtari, A.} et~al.(2021){Vehtari, A.}, {Gelman, A.}, {Simpson, D.},
  {Carpenter, B.}, and {Bürkner, P.C.}}]{aki_vehtari_rank-normalization_2021}
\bibinfo{author}{{Vehtari, A.}}, \bibinfo{author}{{Gelman, A.}},
  \bibinfo{author}{{Simpson, D.}}, \bibinfo{author}{{Carpenter, B.}},
  \bibinfo{author}{{Bürkner, P.C.}},
\newblock \bibinfo{title}{Rank-{Normalization}, {Folding}, and {Localization}:
  {An} {Improved} \${\textbackslash}widehat\{{R}\}\$ for {Assessing}
  {Convergence} of {MCMC} (with {Discussion})},
\newblock \bibinfo{journal}{Bayesian Analysis} \bibinfo{volume}{16}
  (\bibinfo{year}{2021}) \bibinfo{pages}{667--718}. \URLprefix
  \url{https://doi.org/10.1214/20-BA1221}. \DOIprefix\doi{10.1214/20-BA1221}.
%Type = Misc
\bibitem[{Su and Yajima(2020)}]{su_r2jags_2020}
\bibinfo{author}{Y.-S. Su}, \bibinfo{author}{M.~Yajima},
  \bibinfo{title}{R2jags: {Using} {R} to {Run} '{JAGS}'},
  \bibinfo{howpublished}{\url{https://CRAN.R-project.org/package=R2jags}},
  \bibinfo{year}{2020}. \URLprefix
  \url{https://CRAN.R-project.org/package=R2jags}, \bibinfo{note}{accessed
  2020-07-09}.
%Type = Article
\bibitem[{{Abdalla, S.M.} and {Galea, S.}(2022)}]{abdalla}
\bibinfo{author}{{Abdalla, S.M.}}, \bibinfo{author}{{Galea, S.}},
\newblock \bibinfo{title}{Invited commentary: Toward a better understanding of
  disparities in overdose mortality},
\newblock \bibinfo{journal}{American Journal of Epidemiology}
  \bibinfo{volume}{191} (\bibinfo{year}{2022}) \bibinfo{pages}{1280--1282}.
%Type = Article
\bibitem[{{Sumetksy, N.} et~al.(2021){Sumetksy, N.}, {Mair, C.},
  {Wheeler-Martin, K.}, {Magdalena, C.}, {Waller, L.A.}, {Ponicki, W.R.}, and
  {Gruenewald, P.J.}}]{sumetsky}
\bibinfo{author}{{Sumetksy, N.}}, \bibinfo{author}{{Mair, C.}},
  \bibinfo{author}{{Wheeler-Martin, K.}}, \bibinfo{author}{{Magdalena, C.}},
  \bibinfo{author}{{Waller, L.A.}}, \bibinfo{author}{{Ponicki, W.R.}},
  \bibinfo{author}{{Gruenewald, P.J.}},
\newblock \bibinfo{title}{Predicting the future course of opioid overdose
  mortality: An example from two us states},
\newblock \bibinfo{journal}{Epidemiology} \bibinfo{volume}{32}
  (\bibinfo{year}{2021}) \bibinfo{pages}{61--69}.
%Type = Misc
\bibitem[{{Georgia Department of Public Health (GADPH)}(2021)}]{gadph}
\bibinfo{author}{{Georgia Department of Public Health (GADPH)}},
  \bibinfo{title}{Drug surveillance unit: Drug overdose-mortality web query},
  \bibinfo{howpublished}{\url{https://www2.census.gov/programs-surveys/popest/technical-documentation/methodology/2010-2019/natstcopr-methv2.pdf}},
  \bibinfo{year}{2021}. \URLprefix
  \url{https://oasis.ga.us/oasis/webquery/qryDrugOverdose.aspx.},
  \bibinfo{note}{accessed on 09-01-2022}.
%Type = Article
\bibitem[{{Krieger, N.} et~al.(2016{\natexlab{a}}){Krieger, N.}, {Waterman,
  P.D.}, {Spasojevic, J.}, {Li, W.}, {Maduro, G.}, and {Gretchen,
  V.W.}}]{krieger}
\bibinfo{author}{{Krieger, N.}}, \bibinfo{author}{{Waterman, P.D.}},
  \bibinfo{author}{{Spasojevic, J.}}, \bibinfo{author}{{Li, W.}},
  \bibinfo{author}{{Maduro, G.}}, \bibinfo{author}{{Gretchen, V.W.}},
\newblock \bibinfo{title}{Public health monitoring of privilege and deprivation
  with the index of concetration at the extremes},
\newblock \bibinfo{journal}{American Journal of Public Health}
  \bibinfo{volume}{106} (\bibinfo{year}{2016}{\natexlab{a}})
  \bibinfo{pages}{256--263}.
%Type = Article
\bibitem[{{Krieger, N.} et~al.(2016{\natexlab{b}}){Krieger, N.}, {Waterman,
  P.D.}, {Spasojevic, J.}, {Li, W.}, {Maduro, G.}, and {Van Wye, G.}}]{ice1}
\bibinfo{author}{{Krieger, N.}}, \bibinfo{author}{{Waterman, P.D.}},
  \bibinfo{author}{{Spasojevic, J.}}, \bibinfo{author}{{Li, W.}},
  \bibinfo{author}{{Maduro, G.}}, \bibinfo{author}{{Van Wye, G.}},
\newblock \bibinfo{title}{Public health monitoring of privilege and deprivation
  using the index of concentration at the extremes (ice)},
\newblock \bibinfo{journal}{American Journal of Public Health}
  \bibinfo{volume}{106} (\bibinfo{year}{2016}{\natexlab{b}})
  \bibinfo{pages}{253––265}. \DOIprefix\doi{10.2105/AJPH.2015/302955}.
%Type = Article
\bibitem[{{Krieger, N.} et~al.(2016{\natexlab{c}}){Krieger, N.}, {Kim, R.},
  {Feldman, J.}, and {Waterman, P.D.}}]{ice2}
\bibinfo{author}{{Krieger, N.}}, \bibinfo{author}{{Kim, R.}},
  \bibinfo{author}{{Feldman, J.}}, \bibinfo{author}{{Waterman, P.D.}},
\newblock \bibinfo{title}{Using the index of concentration at the extremes at
  multiple geographic levels to monitor health inequities in an era of growing
  spatial social polarization: Massachusetts, usa (2010-2014)},
\newblock \bibinfo{journal}{International Journal of Epidemiology}
  \bibinfo{volume}{47} (\bibinfo{year}{2016}{\natexlab{c}})
  \bibinfo{pages}{788--819}. \DOIprefix\doi{10.2105/AJPH.2015/302955}.

\end{thebibliography}
 \bibliographystyle{elsarticle-num-names}

\clearpage

 \begin{appendix}
 \section{Summary of the Berkson Error Model}\label{sec:berkson}
 
 The Berkson error model (Berkson, 1950)\citet*{raymond_j_carroll_measurement_2006, gustafson} sets the true value $X_i$ equal to the estimated error-prone value $W_i$ plus a measurement error term $U_i$,
\[X_i = W_i + U_i\]
where $E(U_i|W_i) =0$ so that the true value has more variability than the estimated value. There is an interesting relationship at a technical level between error models and regression calibration, where a for $\bm{X}$ given $\bm{W}$ is needed, but we start with a model for $\bm{W}$ given $\bm{X}$. If one has a strucutral model so that one knows the marginal distribution of $\bm{X}$, then an error model can be converted into a regression calibration model by Bayes theorem. Specifically,
\begin{equation}
    f_{\bm{X|W}}(x|w) =\frac{f_{\bm{W|X}} (w|x) f_{\bm{X}}(x)}{\int f_{\bm{W|X}} (w|x) f_{\bm{X}} dx},
\end{equation}
where $f_{\bm{X}}$ is the density of $\bm{X}$, $f_{\bm{W|X}}$ is the density of $\bm{W}$ given $\bm{X}$, and $f_{\bm{X|W}}$ is the density of $\bm{X}$ given $\bm{W}$. For example, suppose that $\bm{W}= \bm{X} + \bm{U}$, where $\bm{X}$ and $\bm{U}$ are uncorrelated. As such, the best linear predictor of $\bm{X}$ given $\bm{W}$ is $(1-\lambda)E(\bm{X}) + \lambda \bm{W}$, and
\begin{equation}\label{eq:xx}
    \bm{X} = (1-\lambda) E(\bm{X}) + \lambda \bm{W} + \bm{U^*}
\end{equation}
where $\lambda = \sigma^2_x /(\sigma^2_x + \sigma^2_u)$ is the attenuation, $\bm{U^*} = (1-\lambda) \{\bm{X} - E(\bm{X}) - \lambda \bm{U}$. Equation \ref{eq:xx} has the form of a Berkson model, even though the error model is classical. Note, however, that the slope of $\bm{X}$ on $\bm{W}$ is $\lambda$. Therefore, the variance of $\bm{X}$ is smaller than the variance of $\bm{W}$ in keeping with the classical rather than Berkson errors.

\section{Mapped relative error}\label{sec:relerr}
\begin{figure}[H]
\center
\includegraphics[width=\textwidth]{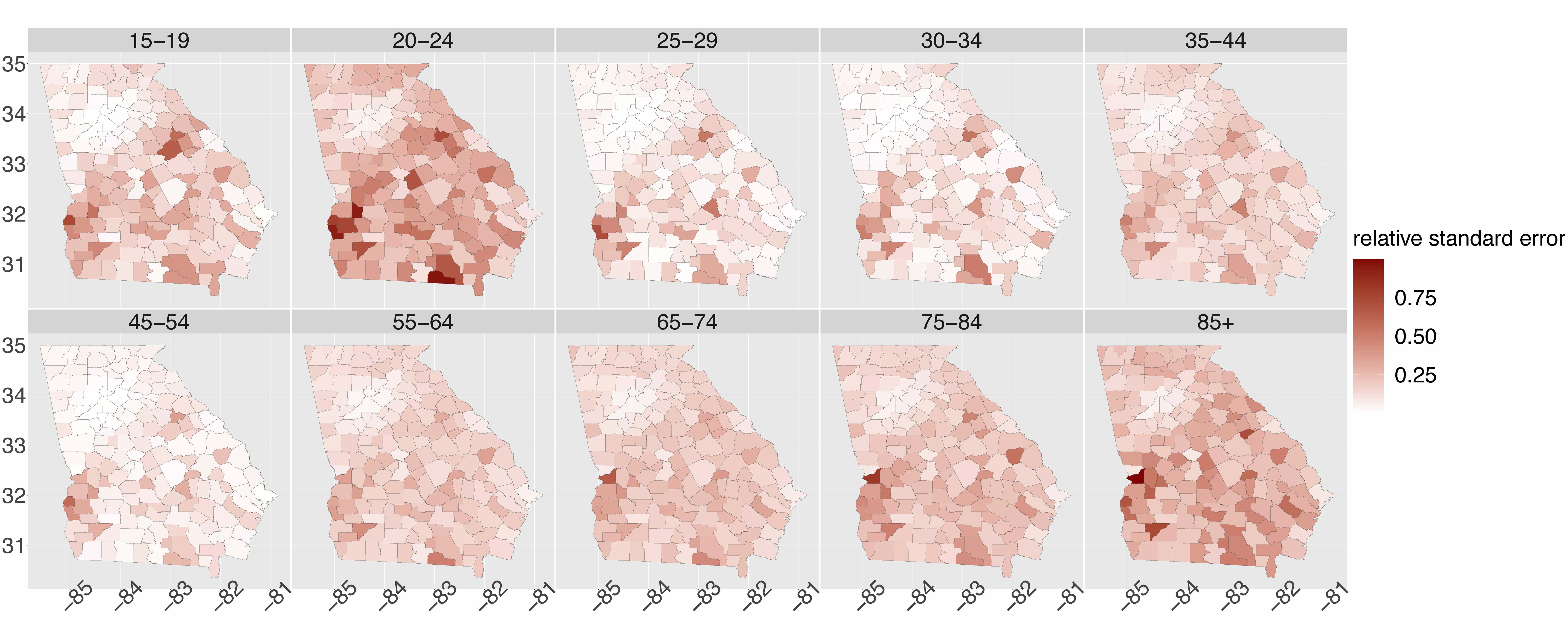}
\caption{Mapped ACS reported relative errors across 159 counties in Georgia, stratified by 5 year age group intervals. Scale ranges from small offset relative error (white) to large offset relative error (dark red).}
\label{fig:acsrelerr}
\end{figure}

Figure \ref{fig:acsrelerr} maps the relative error defined as the ACS reported standard error divided by the population size. By mapping relative errors across 159 counties in Georgia, we account for the error relative to the size of the population. Figure \ref{fig:acsrelerr} illustrates that counties with larger population sizes and more heterogeneity in the population have smaller relative errors compared to smaller population counties. 
 
\section{Description of the ICE Index}\label{sec:ice}
The Index of Concentration at Extremes (ICE) is a measure used to capture economic polarization between White and Black residents within an area. It is derived based on ACS reported variables, given by:
\[ICE_{i} = (W_{i} - B_{i})/n_{i}\]
in which $W_{i}$ denotes the ACS reported number of affluent White households in area $i$, i.e., the top $5^{th}$ percent of all White household incomes. ACS reported number of poor Black households, in area $i$, is denoted $B_{i}$, i.e., the bottom $5^{th}$ percent of all Black household incomes. To determine economic disparity across White and Black households, the ICE measures takes the ratio of the difference between the number of affluent White households and the number of poor Black households over the total number of households, denoted $n_{i}$. An ICE measure of 1 denotes that all households are in the White privileged group, in contrast, -1 denotes all households are in the Black deprived group \citep{ice1, ice2}.

\section{Graphical depiction of covariate relationships}\label{sec:covs}
\begin{figure}[H]
\center
\includegraphics[width=0.8\textwidth]{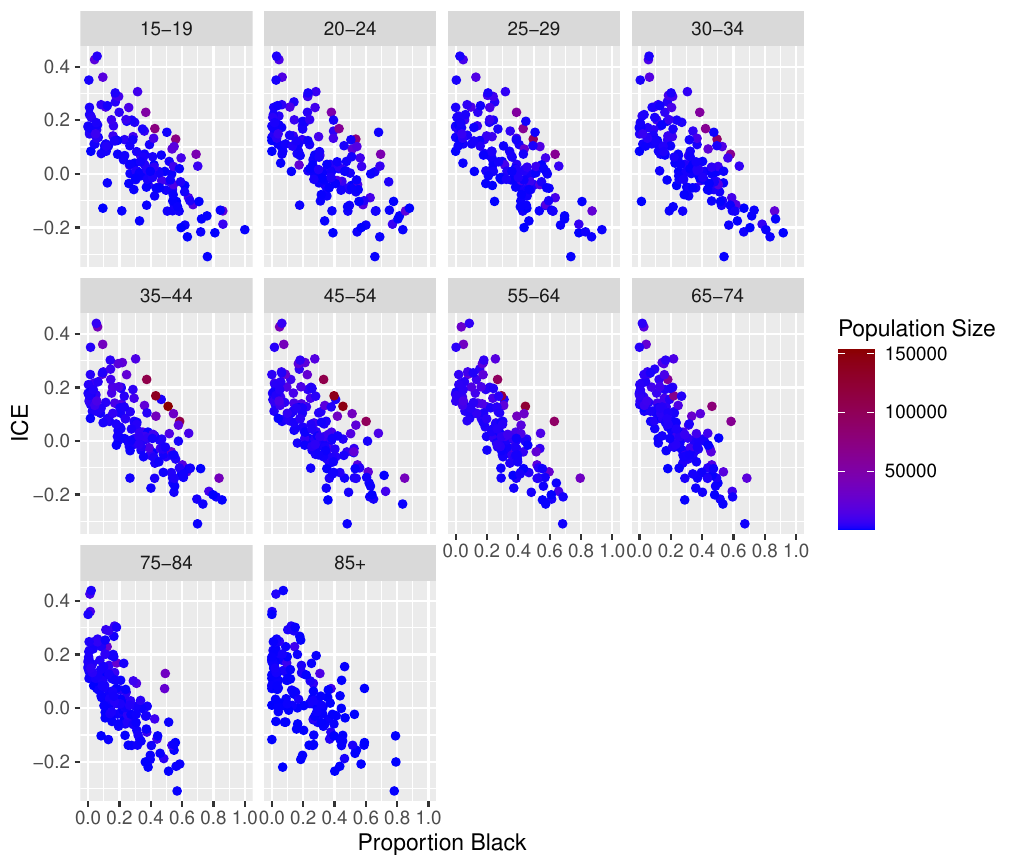}
\caption{Scatterplot of the relationship between the proportion Black population (x-axis) and the ICE measure (y-axis) where color indicates the size of the population (blue to red indicates smaller to larger sizes). We illustrate the relationship separately for each age-group. }
\label{fig:covs}
\end{figure}

 Figure \ref{fig:covs} illustrates a graphical representation of the relationship between the proportion Black population (i.e., the proportion of the population that racially identify as Black) by county and the county-specific ICE measure broken down by age-group. Additionally, we indicate larger population sizes in red, and smaller population sizes in blue. There is a notable negative trend across all age-groups in which a higher proportion Black population in a given county is associated with lower ICE values. Conversely, counties with lower proportions of Black populations are associated with higher ICE measures. There is no obvious trend by population size. This trend across age groups indicates that for counties with smaller Black proportions of their populations (e.g., less than 0.4), ICE measures indicate increased numbers of households in the White privilege group which increases as this proportion gets smaller. Conversely, as the Black proportion of the population increases, there is a moderate trend towards increased numbers of households in the Black deprived group.

\bigskip

\section{Mapped County Estimates for Georgia}\label{sec:all}
\includepdf[pages=-]{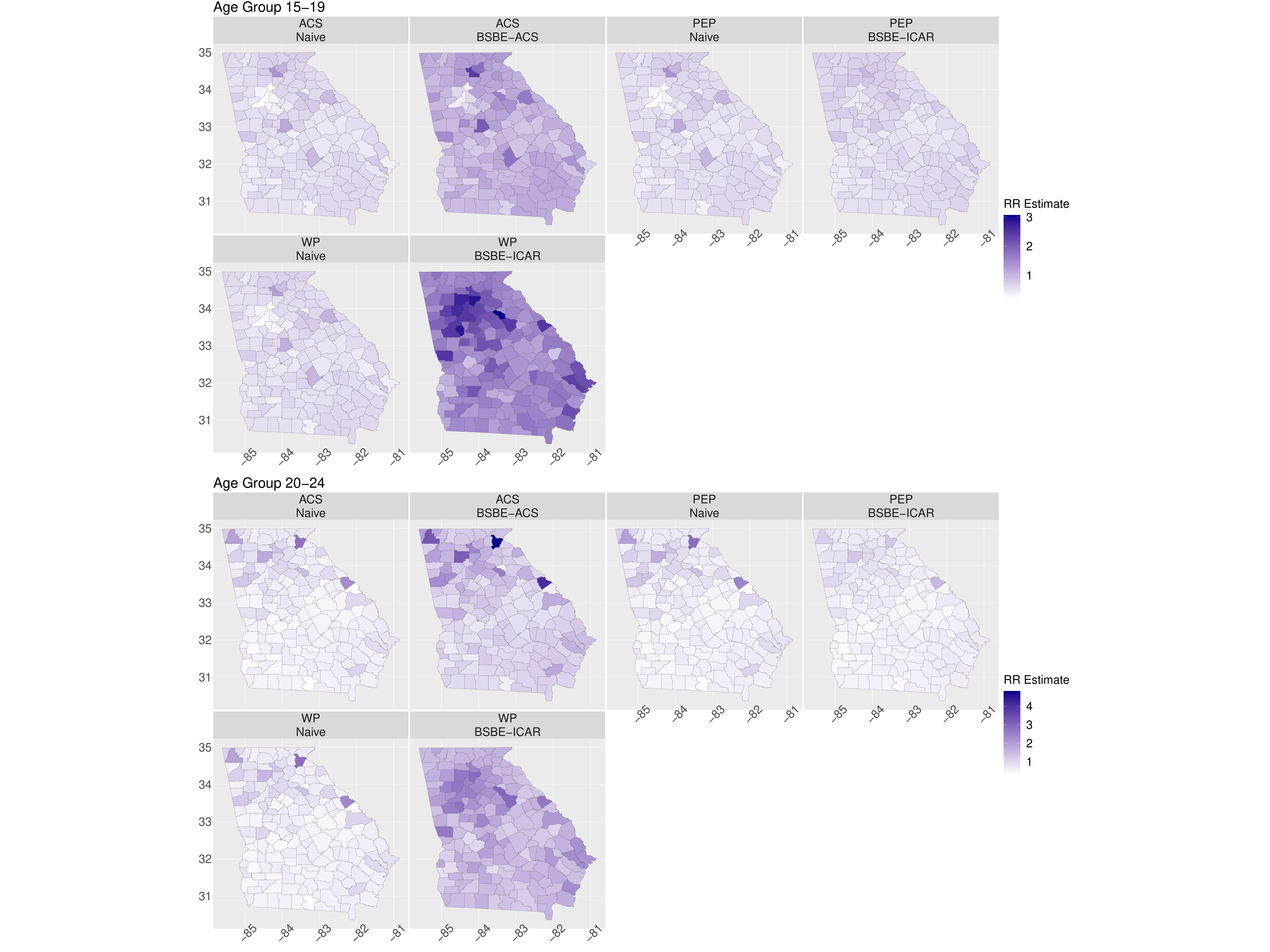}
\end{appendix}

 \end{document}